\documentclass[%
 reprint,
 amsmath,amssymb,
 aps,
]{revtex4-2}

\usepackage[utf8]{inputenc}
\usepackage{bm}
\usepackage{floatrow}
\usepackage{soul}

\usepackage{amsmath}
\usepackage{amssymb}
\usepackage{slashed}
\usepackage{mathcomp}
\usepackage{physics}
\usepackage{latexsym}
\usepackage{cancel}
\usepackage{isotope}
\usepackage{comment}

\makeatletter
\renewcommand\@make@capt@title[2]{%
 \@ifx@empty\float@link{\@firstofone}{\expandafter\href\expandafter{\float@link}}%
  {\textbf{#1}}\@caption@fignum@sep#2\quad
}%
\makeatother

\usepackage{graphicx} 

\begin{document}

\preprint{APS/123-QED}

\title{A new technique to measure gravitational mass of ultra-cold matter and its implications for antimatter studies}

\author{Boaz Raz}
\email{raz.boaz@weizmann.ac.il}
\author{Gavriel Fleurov}
\email{gavriel.fleurov@weizmann.ac.il}
\author{Roi Holtzman}
\author{Nir Davidson}
\affiliation{Department of Physics of Complex Systems, Weizmann Institute of Science, Rehovot 7610001, Israel}
\author{Eli Sarid}
\affiliation{Department of Physics, Ben-Gurion University of the Negev, P.O.B. 653, Beer-Sheva 84105, Israel}
\date{\today}

\begin{abstract}
Measuring the effect of gravity on antimatter is a longstanding problem in physics that has significant implications for our understanding of the fundamental nature of the universe. Here, we present a technique to measure the gravitational mass of atoms, motivated by a recent measurement of antimatter atoms in CERN \cite{anderson_observation_2023}. We demonstrate the results on ultra-cold atoms by measuring the surviving fraction of atoms gradually released from a quadrupole magnetic trap, which is tilted due to gravitational potential. We compare our measurements with a Monte Carlo simulation to extract the value of the gravitational constant. The difference between the literature value for $g$, the local acceleration due to gravity, and the measured value is $(-1.9 \pm 12^{stat} \pm 5^{syst}) \times 10^{- 4} g$. We demonstrate the importance of various design parameters in the experiment setup, and estimate their contribution to the achievable accuracy in future experiments. Our method demonstrates simplicity, precision, and reliability, paving the way for future precision studies of the gravitational force on antimatter. It also enables a precise calibration of atom traps based on the known gravitational attraction of normal matter to Earth.
\end{abstract}

\maketitle

\subsection{Introduction}

The study of antimatter is an active area of research in physics \cite{tino_precision_2020}. According to General Relativity (GR), matter and antimatter should have the same gravitational free-fall due to the Weak Equivalence Principle (WEP) \cite{chardin_gravity_2018}. Recent theories suggest that antimatter might have a different free-fall acceleration than matter \cite{indelicato_gbar_2014}. A violation of the WEP will lead to physics beyond GR. However, measuring the effect of the gravitational force on antimatter has been challenging, mainly due to the extreme difficulty of producing, trapping, and manipulating antimatter in sufficient quantities for experimental study.

Recently, the ALPHA collaboration in CERN has demonstrated for the first time a measurement of the free fall of antimatter atoms in Earth's gravitational field \cite{anderson_observation_2023}. The result is consistent with a fall similar to the fall of normal matter atoms. This result opens the way to more precise measurements that will characterize the gravitational interaction between neutral antihydrogen atoms and Earth's gravitational field and provide a more stringent test of the WEP. Such precise determination depends on fully understanding systematic measurement errors and perfecting the simulations that describe the experiments. Hopefully, The next generation of experiments will attempt to determine the gravitational mass interaction of antimatter atoms with $1\%$ uncertainty \cite{bertsche_prospects_2018}.

Several methods measure gravitational acceleration $g$ on atomic matter with high accuracy. These include 10-meter high atomic fountains with $10^{-12} g$ \cite{dickerson_multiaxis_2013} or more compact atom interferometers with long interrogation time \cite{xu_probing_2019}, which produce an interference pattern sensitive to gravitational potential differences. However, applying such techniques to antimatter is difficult as they require large quantities of ultra-cold atoms.

\begin{figure}[ht]
\centering
    \includegraphics[width=\textwidth]{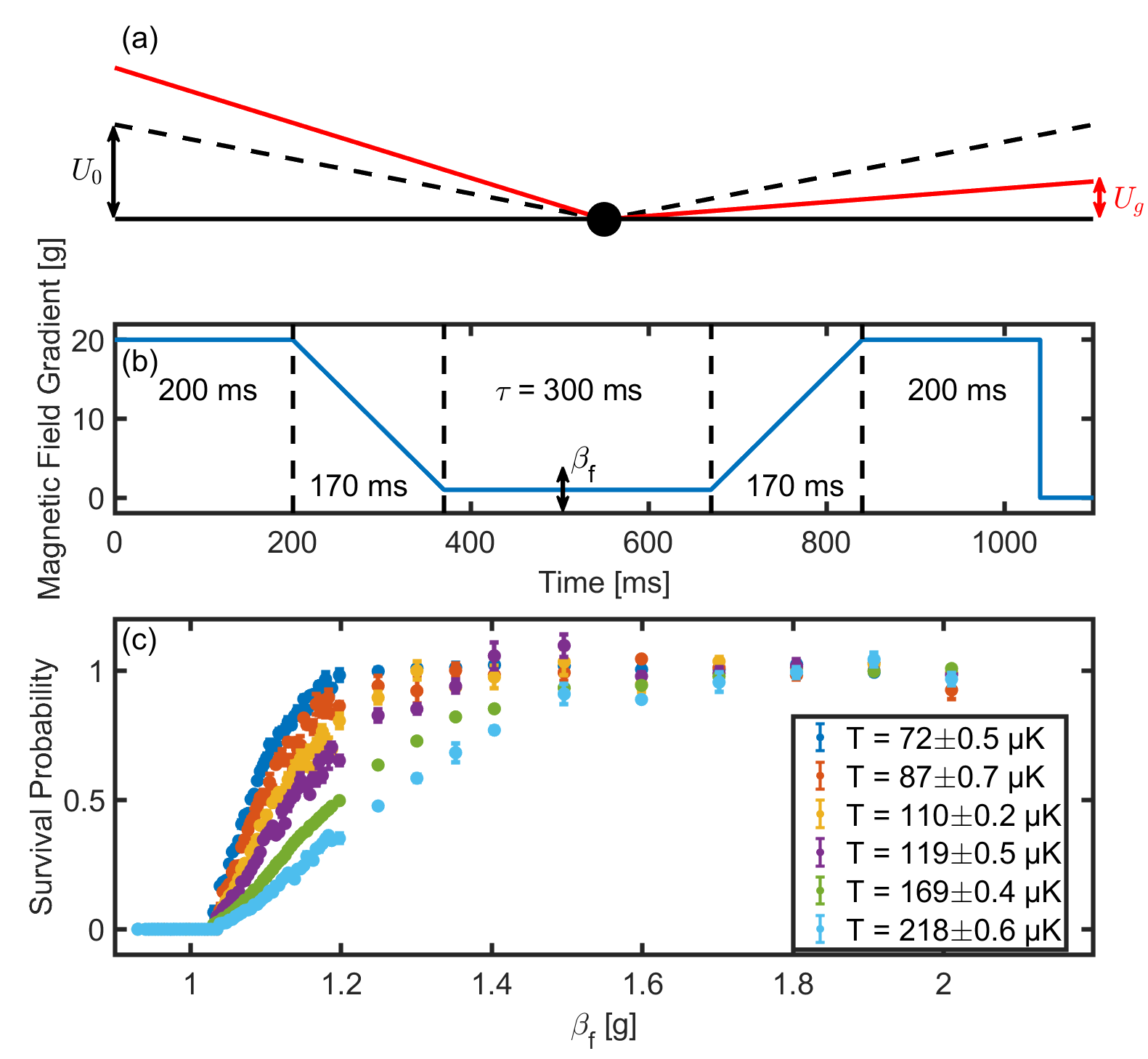}
    \caption{The setup, protocol, and results. (a)  Illustration of atoms (black dot) in a tilted quadrupole magnetic trap (red) due to gravitational potential. The black dashed line is the magnetic field potential without gravity. Trap depth with and without gravity at the bottom cell wall is given by $U_g$ and $U_0$, respectively. The ratio between the magnetic field gradient and gravity determines the tilting angle. (b) The experimental protocol: magnetic field gradient (divided by $\beta_*$) as a function of time. Atoms are prepared in a strong trap ($\beta_i$), followed by an adiabatic linear decrease to varying values $\beta_f$, where they are held for time $\tau$ before adiabatically increasing back to $\beta_i$. The atoms are then released for absorption imaging after $4$ ms of expansion. (c) Measured survival probability of trapped atoms as a function of $\beta_f$ for various temperatures. The loss features exhibit narrower profiles at lower temperatures.}
    \label{Fig:1}
\end{figure}

In this paper, we introduce a technique to measure the gravitational interaction of atoms analogous to the technique ALPHA applies with antihydrogen atoms and experimentally validate it using ordinary matter. As in ALPHA, our approach involves the gradual release of ultra-cold atoms in a magnetic trap, which is biased due to the influence of gravity. 
We use the simple quadrupole magnetic trap \cite{almog_high_2018,stuhler_continuous_2001} and precisely adjust its constant and well calibrated magnetic field gradients until the gravitational and magnetic potentials perfectly counterbalance each other (see Fig. Fig. \ref{Fig:1} (a)), causing a sharp and easily interpreted loss feature. Subsequently, we quantify the residual fraction of atoms within the trap and compare our findings with simulations to extract $g$. The difference between our extracted value of $g$ and the established value is $(-1.9 \pm 12^{stat} \pm 5^{syst}) \times 10^{- 4} g$. 

Our method is simple, easy to calibrate and model \cite{raz_see_2023}, and can yield high precision measurements by properly choosing experiment parameters. While there are obvious differences between our experiment and the antihydrogen experiment, the technique we developed is useful for studying systematic errors and analysis methods employed with antimatter experiments. Because of the relative ease of performing such experiments with normal atoms, parametric studies can be performed with a high degree of control of parameters such as the temperature and number of atoms.

Our study also offers insights relevant to the broader domain of cold atom physics. These insights are especially useful when implementing adiabatic cooling and truncating the Boltzmann distribution of atoms within an optical tweezer \cite{tuchendler_energy_2008}. Our work on the survival probability of trapped atoms can contribute to designing more efficient protocols for preparing atoms at low microkelvin temperatures. Finally, our protocol can accurately calibrate normal matter traps using the known value of the gravitation field.

\subsection{Experiment}

\begin{figure}[ht]
    \centering
    \includegraphics[width=\textwidth]{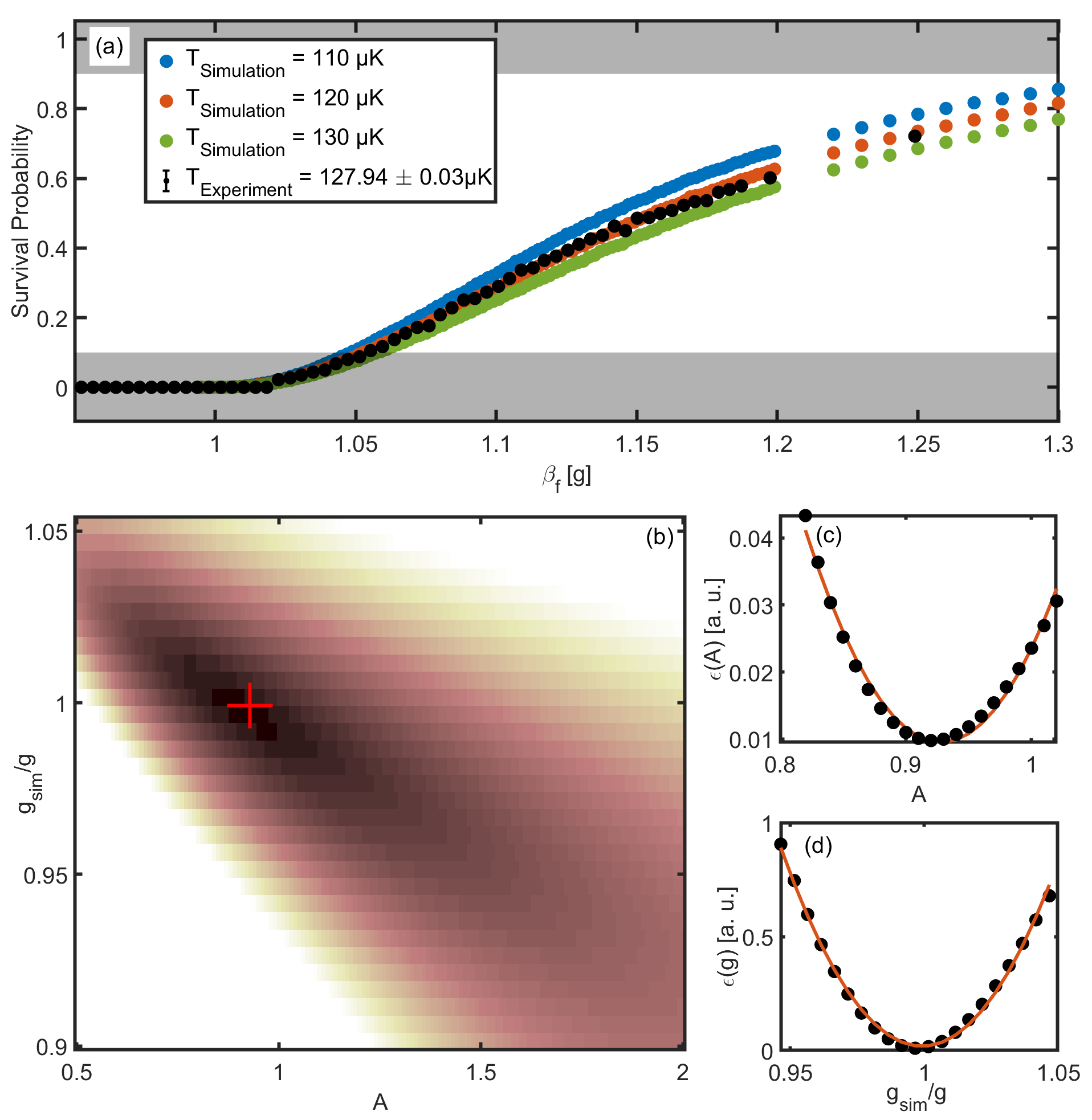}
    \caption{A comparison of experimental data and simulation. (a) Experimental data (black) agrees with simulation data at similar temperatures. The range of analysis (10\%-90\%) is shown between the gray-shaded areas. (b) The error function $\epsilon(g,A)$ is plotted as a function of $A$ and $g$. The red cross indicates the global minimum. (c) and (d) Parabola fits cross-sections at the minimum error surface along $A$ and $g$, respectively.}
    \label{Fig:2}
\end{figure}

The motion of the atoms in a quadrupole magnetic trap and under the influence of a gravitational field is determined by the potential \cite{almog_high_2018,stuhler_continuous_2001}
\begin{align}
    U(x,y,z) & = g_F m_F \mu_B \beta \sqrt{\frac{x^2+y^2}{4}+z^2} + m g z,    
    \label{eq:EOMs}
\end{align}

\noindent where $g_F$ is the hyperfine Lande g-factor, $m_F$ is the magnetic quantum number of the hyperfine level $F$, $\mu_B$ is Bohr's magneton, $\beta$ is the magnetic field gradient in the $z$ direction, which we choose to point opposite to the gravity, $g = 9.794608\ \text{m}/\text{s}^2$ is the gravitational acceleration in our lab \cite{bislin_earth_2018} and $m$ is the atomic mass. This potential is depicted in Fig. \ref{Fig:1}(a) along the $z$-axis ($x = y = 0$); there, one sees that the two linear forces are acting in the same direction ($z>0$) or in opposite directions ($z< 0$). Thus, we compare the coefficients of both terms and define the normalization factor of the magnetic field gradient ($\beta$) axis, $\beta_* = m g /g_F m_F \mu_B$. At $\beta = \beta_*$, the potential for $z<0$ (red line in Fig. \ref{Fig:1}(a)) is exactly zero. In the case of using atomic properties of antimatter, the uncertainties associated with these properties are generally higher when compared to the well-known values for regular matter. This is primarily due to the experimental challenges and limited data available for antimatter systems. However, owing to recent developments in the anti-hydrogen field \cite{ahmadi_observation_2018}, our method will be mostly limited by the accuracy of magnetic field calibration.

In our experiment, we load between $5 \times 10^6$ to $50 \times 10^6$ \isotope[87]{Rb} atoms in the low-field seeking magnetic state $\ket{F=2,m_F=2}$ into a quadrupole magnetic trap \cite{petrich_stable_1995, davis_evaporative_1995,fang_cross-dimensional_2020,stuhler_continuous_2001}. Using forced Radio-Frequency (RF) evaporation \cite{davis_evaporative_1995,luiten_kinetic_1996,ketterle_boseeinstein_1996, muller_bose-einstein_2000,cohen-tannoudji_advances_2011}, we cool the atoms to temperatures $T$ ranging between $70-230 \mu K$. The limiting factor to working with lower temperatures is losses due to Majorana spin-flips \cite{petrich_stable_1995, davis_evaporative_1995,sukumar_spin-flip_1997,majorana_ettore_2006,lin_rapid_2009}, where the magnetic quantum state of the atom can change to an untrappable state due to a large rate of change in the magnetic field $\abs{\flatfrac{\Dot{B}}{B}}$ compared to the Larmor frequency $\omega_L$ violating the adiabatic condition $\abs{\flatfrac{\Dot{B}}{B}} \ll \omega_L$. At lower temperatures, this effect becomes dominant, substantially reducing the lifetime in the trap and distorting the measured loss feature \cite{raz_see_2023}. 

Following the experimental protocol depicted in Fig. \ref{Fig:1}(b), we decrease the trap strength to various final values $\beta_f$ and hold the atoms for $\tau = 300$ ms. The change in the field is slow enough to allow the temperature of the cloud to follow the magnetic field adiabatically $(T\propto \beta ^{2/3})$ \cite{almog_high_2018, raz_see_2023}. Subsequently, we adiabatically restore the trap to its initial strength to avoid systematic errors during imaging. We then release and count the atoms using absorption imaging after 4 ms of expansion \cite{anderson_observation_1995,ketterle_making_1999}. The distance from the center of the quadrupole trap to the walls is $L = 2.5 \pm 0.1$ mm; atoms reaching the walls are lost and do not appear in the image.

The experiment was conducted across 16 different temperatures. Due to limitations in the standard time-of-flight method \cite{anderson_observation_1995,ketterle_making_1999}, such as unintentional adiabatic cooling during the shutdown of the magnetic trap and eddy currents distorting the atom cloud shape, accurate temperature measurement is challenging. Therefore, we extract the atomic temperature by comparing the measured and simulated fraction of surviving atoms. We verify that these extracted temperatures are consistent with the less accurate time-of-flight method, as shown in Fig. 10 in \cite{raz_see_2023}. In this paper, 'temperature' refers to the value obtained from the simulation. This value is also the one utilized in our analysis to determine the measured value of the gravitational constant $(g)$.

To ensure the stability and accuracy of the measurement, the magnetic field gradients are generated by driving the electric currents using a custom-built stabilized current supply with $2$ ppm stability and noise. We use a hall probe (\href{https://www.lem.com/en/product-list/it-200s-ultrastab}{LEM IT 200-S ultrastab}) with $1$ ppm/month stability that passes currents of up to $30$ mA through a $50 \Omega$ resistor (Vishay VHP4ZT) with $0.05$ ppm/K stability. The ambient temperature in the lab is stabilized to $0.1$ K. We then measure the resulting voltage with a Digital Multimeter with a $16$ ppm precision (\href{https://www.keysight.com/il/en/product/34470A/digital-multimeter-7-5-digit-truevolt-dmm.html}{Keysight TrueVolt 34470A}) calibrated and measured for it's offset value \cite{raz_see_2023}. By careful calibration, \cite{raz_see_2023}, we can determine the magnetic field gradient on the atoms ($\beta_f$) from accurate measurements of the current in the magnetic field coils in each sequence after $\tau/2$.

\subsection{Results and Analysis}

Figure \ref{Fig:1}(c) shows the normalized  \cite{raz_see_2023} survival probability as a function of $\beta_f$ measured at six representative temperatures. We observe a temperature-dependent loss feature for $g <\beta_f $. As expected, hotter temperatures lead to a broader loss feature. Conversely, colder temperatures lead to distorted loss features due to the effect of Majorana spin-flips. In the experiment, a small fraction of atoms are in state $\ket{F=2,m_F=1}$ due to imperfect initial state preparation. These atoms also have a $\beta$ dependant loss feature, where all atoms are lost at $\beta_f = 2 g $ because their magnetic moment is twice weaker than $\ket{F=2,m_F=2}$, creating a distortion of the loss feature of the $\ket{F=2,m_F=2}$ atoms, thus limiting the range of $\beta_f$ used in the experiment.

We compare our experimental measurements to Monte Carlo simulations \cite{raz_see_2023}, as depicted in Fig.\ref{Fig:2}(a). The simulation propagates $10^5$ atoms, starting from an initial equilibrium distribution  $n(x,y,z) = n_0\exp(- U(x,y,z) /k_B T)$ \cite{stuhler_continuous_2001}, and using the equations of motion derived from Eq. (\ref{eq:EOMs}); here, $n_0$ is the peak density and $k_B$ is the Boltzmann constant. To extract the  gravitational acceleration $g$, we minimize the distance between the measurement and simulation given by the error function
\begin{align}
    \epsilon(g,T,A) = \sum_j \frac{\qty(A \times P_j - f(\beta_{f,j};T,g))^2}{\sigma_{P,j}^2}
\end{align}
where $P_j$ is the measured survival probability, $f(\beta_{f,j}; T,g)$ is the simulated survival probability, $\beta_{f,j}$ are the final magnetic gradient values of the experiment, and $\sigma_{P,j}$ is the one standard deviation uncertainty in the measured value of $P_j$. We only use the loss feature's $10\% - 90\%$ interval to avoid noisy measurements at low numbers of atoms and fitting bias to the plateau. We use $A$, a normalization factor, as a fit parameter due to limitations on extracting the initial number of atoms from our measured loss features  \cite{raz_see_2023}: at low temperatures, Majorana losses, not included in the simulations,  distort the plateau; at higher temperatures, the plateau is reached only at $\beta_f > 2$ where it is distorted by the contribution of atoms at $\ket{F=2,m_F=1}$. 

\begin{table}[ht]
    \caption{Error budget of one specific measurement set at $T = 127.94 \pm 0.03 \mu K$}
    \centering
    \begin{tabular}{c|p{1.2cm}|p{1.2cm}|p{1.3cm}|p{1.2cm}}
        Source             & Stat. [abs.]  & Stat. [$10^{-4}g$]& Syst. [abs.]& Syst. [$10^{-4}g$]\\ \hline \hline                 
        Fit                & 0.02\%&  2&                &\\           \hline 
        Field  Calibration & 0.05\%        &  5                   & 0.05\%         & 5  \\ \hline        
        Current            & 2 ppm         &  0.02                &  16 ppm        & 0.16\\  \hline
        Temperature        & 0.02 \%&  0.25& & \\ \hline
        Normalization      & 0.23 \%& 2.4&               &\\ \hline      
        \textbf{Total}     &               & \textbf{5.4}&               & \textbf{5}\end{tabular}
    \label{tab:error table}
\end{table}

We run the simulation for a range of values of the gravitational acceleration $g$,  normalization factor $A$, and atomic temperate $T$. We find the position of the global minima of the error function $\epsilon(g,T,A)$, which is smooth and convex. To extract the values of the parameters $(g,T,A)$ from the corresponding cross-sections around the global minima, as depicted in Fig. \ref{Fig:2}(b-d), we use a parabolic fit of the following form

\begin{equation}
    f = M(g - g_0)^2 + C
    \label{eq:parabola_fit_func}
\end{equation}

Where $g_0$, the extracted value for the measured gravity, is replaced by $T_0,A_0$ when extracting the temperature and the normalization factor, respectively. 

Various sources of uncertainty in our measurements impact the determination of $g$. Any uncertainty or error in the calibration of the magnetic field gradient results in an inaccurate measurement of $\beta_*$ and an incorrect estimation of $g$. Uncertainty in the number of atoms is mitigated by using randomized measurements and fitting the normalization factor $A$. We use simulations to estimate how the uncertainty in $A$ propagates through $\pdv{g}{A}$. Temperature errors and uncertainties propagate similarly \cite{raz_see_2023}. The combined fitting and experimental uncertainties determine the total uncertainty in $g$.

\begin{align}
    \Delta g^2 = \Delta g_0^2 + \Delta \beta^2 + \left(\frac{\partial g}{\partial T} \right)^2 \Delta T^2 + \left(\frac{\partial g}{\partial A} \right)^2 \Delta A^2,
\end{align}
Where $\Delta g_0$, $\Delta T$ and $\Delta A$ are the $68\%$ confidence interval for the fit parameter of the respective parabola minimum of the extracted gravity $g_0$, temperature $T_0$ and normalization factor $A_0$, as described in Eq. \ref{eq:parabola_fit_func}, $\Delta \beta$ is the total uncertainty in the magnetic field gradient due to uncertainties in field calibrations and current measurements. The derivatives of $g$ with respect to $T$ and $A$ are calculated from the simulation \cite{raz_see_2023} and summarized in Table \ref{tab:error table}. Furthermore, several contributions have systematic errors \cite{raz_see_2023}. In units of $g$, the systematic errors are 0.05 \% in the magnetic field calibration
and 16 ppm in the control of the current in the quadrupole coils.

In our experiment, a broad loss feature arises due to the very short distance (2.5 mm) the atoms fall before hitting the cell wall. Increasing this distance can dramatically narrow the loss feature (see below) to improve the accuracy of the extracted value of $g$ significantly. For a sufficiently narrow loss feature, the limiting factor of our method will be the calibration error of the magnetic field gradient. On the other hand, in our compact system, maintaining and accurately measuring a linear magnetic field across the entire experiment area is manageable, whereas, for larger systems, this can be challenging. Therefore, enhancing our method's sensitivity requires a compromise between achieving a narrow loss feature and accurately determining its position.
 
We compare our method's relative uncertainty and compute the absolute deviation $\delta g = g - g_\text{meas}$. We compute the weighted mean of points taken between $80 $ \textmu K $ < T < 200 $ \textmu K to find $\delta g/g = (-1.9 \pm 12) \times 10^{- 4}$. Refer to Fig. 11 in \cite{raz_see_2023} for the complete data set and to Fig. 12 for the corresponding uncertainties across all measurement sets. For a consistency check, we also compare these results with values obtained using the measured temperatures, where we find $\delta g/g = (-3.7 \pm 8.2) \times 10^{- 4}$ with unknown systematic uncertainty and shift due to the temperature measurement.

Our larger source of uncertainty is the temperature measurement due to its effect on the width of the loss feature. The next largest source of uncertainty is magnetic field calibration, which shifts the entire position of $g_\text{meas}$. We can mitigate the temperature's contribution when choosing the experimental setup's physical parameters to minimize the loss feature's width, as discussed in Section \ref{seq:optimizing}. The magnetic field calibration can be improved between one and two orders of magnitude \cite{raz_see_2023}.

\subsection{Improving Measurement Accuracy}\label{seq:optimizing}

The width of the loss feature (Fig. \ref{Fig:1}(c)) strongly influences the accuracy and uncertainty of our measurements. Narrower loss features improve the precision of our $g$ measurement and reduce sensitivity to temperature variations. Reducing the temperature below $70 \mu K$ would narrow the loss feature but is impaired in our quadrupole magnetic trap due to Majorana spin-flips. Other traps can mitigate this effect but introduce complications such as distorting the atom distribution and inhomogeneous potential gradients. Hence, we focus our discussion on a temperature of $70 \mu K$ and investigate how the width of the loss feature depends on the falling distance to the cell wall $L$ and the hold time $\tau$.

\begin{figure}[ht]
\centering
    \includegraphics[width= \textwidth]{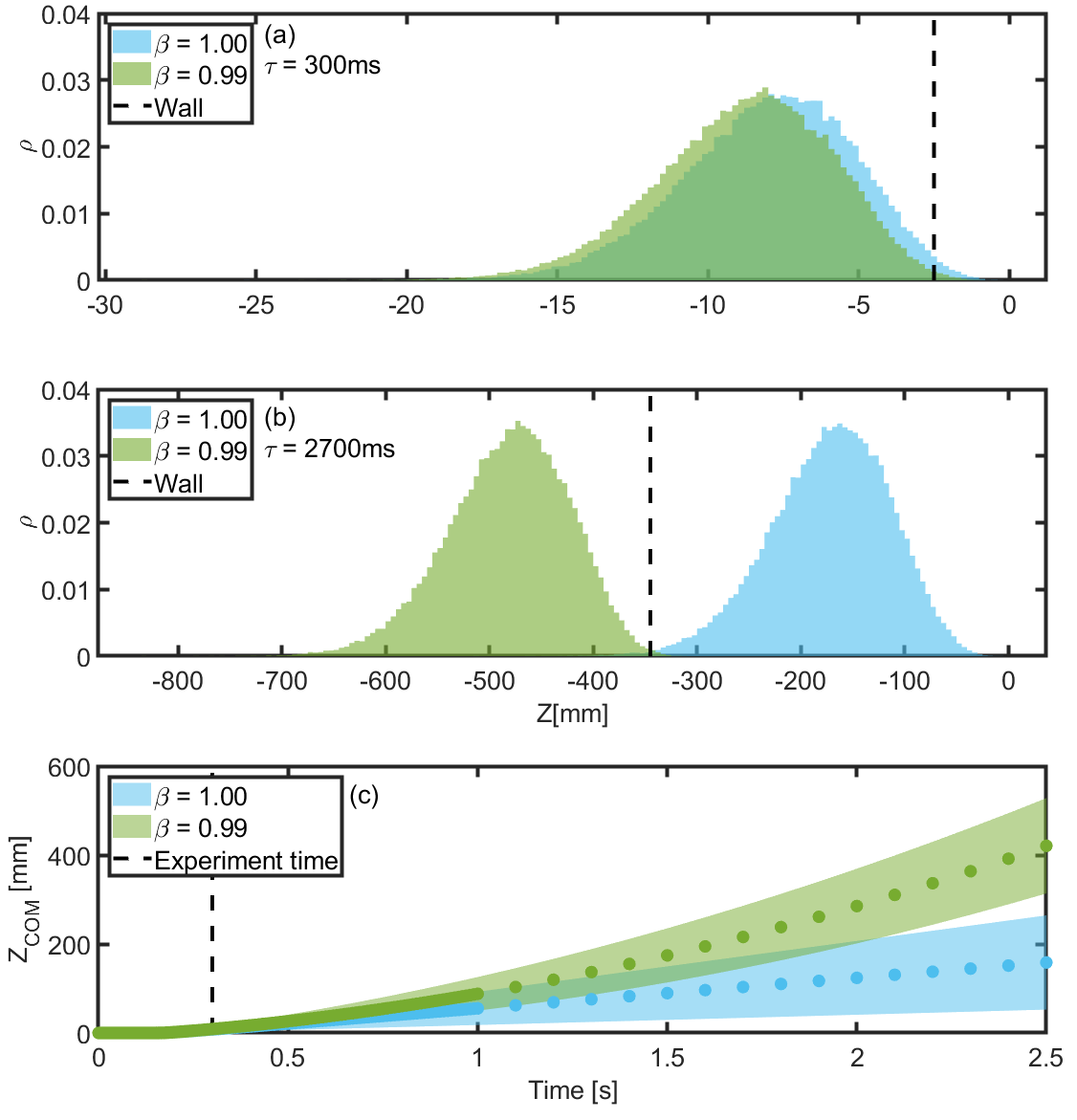}
    \caption{Evolution of simulated density distribution. (a) Density distribution of the Z component of a 3D simulation with one million atoms at $70 \mu K$ after a hold time of $ \tau = 300$ ms, as in the experiment, for different trap strengths. The trap is centered at $Z=0$, and gravity points down along the $Z$ axis. The dashed black line is the position of the cell wall in our experimental setup. The survival probability for each distribution is the area to the right of the wall. (b) Evolution of the same density profiles in (a) after a hold time of $\tau = 2700 $ ms. The dashed line shows the optimal position of the cell wall where almost all atoms are lost for $\beta = 0.99$, whereas almost all atoms survive for $\beta = 1$, making the loss feature extremely narrow. (c) Cloud release dynamics. Center of mass position of the cloud along the $z$-axis for simulations at $T=70 \mu K$ with two different $\beta_f$ on the onset of being completely released from the trap. The shaded area shows the $2\sigma$ width of the simulated distribution. The width of the loss feature corresponds to the overlap between the distributions. The dashed line shows the hold time in the experiment.}
    \label{Fig:3}
\end{figure}

When the axial trapping strength is equal to or weaker than gravity, the vertical density distribution broadens according to the thermal velocity distribution ($v_T = \sqrt{k_B T / m}$), and the center of mass experiences an effective acceleration ($a_{\text{eff}} = g - g_F m_F \mu_B \beta / m$). 
To well resolve between distributions with $\beta_f$ slightly larger and slightly smaller than $g$, the fall due to the $a_{\text{eff}}$ must exceed the broadening of the distribution ($a_{\text{eff}} \tau > v_T$). Therefore, a longer hold time ($\tau$) is essential.

Fig. \ref{Fig:3} illustrates the simulation of the evolution of density distribution, the center of mass, and the expansion of the clouds for  $\beta_f =0.99$ and $\beta_f =1$ after they are released from the trap. As seen in Fig. \ref{Fig:3}a, after $300$ ms, as in our experiment, the clouds are only partly resolved, while after a longer time ($2700$ ms, Fig. \ref{Fig:3}b), they are fully resolved. The improvement of resolution between the clouds with fall time is clearly seen in  Fig. \ref{Fig:3} (c). 

The spatial resolution between the falling clouds can translated into a sharp loss feature around $\beta_f =1$ only when $L$ is properly chosen, as seen in Fig. \ref{Fig:3}: $L=350 $ mm (vertical dashed line in Fig. \ref{Fig:3}b) provides nearly full loss for $\beta_f =0.99$ and nearly no loss for $\beta_f =1$ after $\tau = 2700$ ms, hence an extremely narrow loss feature. Note that an asymmetry between the positions of the upper and lower cell walls can distinguish between downward and upward-falling atoms.

\subsection{Summary and Outlook}

Our work presents a new approach to measuring gravitational acceleration with cold atoms, motivated by the technique applied by the ALPHA collaboration to measure the gravitational free-fall of antihydrogen atoms. By utilizing an available quadrupole magnetic trapping (and not octupole such as in ALPHA) and careful analysis of the loss features, we have demonstrated the potential of this method to measure $g$ with  $\approx 0.1\%$ accuracy, which represents a desirable goal for the accuracy of antimatter gravity experiments. The protocol can be reversed for regular matter, where the precisely known value of gravitational acceleration can be used to calibrate the magnetic field gradient on the cold atoms.

While the temperature in our experiment is substantially lower than the experiment in ALPHA, our work outlines the dependence and importance of lower temperatures. Majorana losses, which are important in our experiment, will not limit the ALPHA experiment, where the temperatures are higher, and the anti-hydrogen atoms are not trapped around the zero of the magnetic field.

While our proof-of-concept experiment has demonstrated the feasibility of this method, further improvements in the experimental setup, such as increasing the vacuum chamber size, can significantly enhance the accuracy of our measurements.  As proposed in Fig. \ref{Fig:3} (b), improvements in system geometry are anticipated to reduce the current error sources by 1-2 orders of magnitude. This would make magnetic field calibration the leading source of error. By incorporating techniques such as microwave spectroscopy for magnetic field gradient calibration, the accuracy of our measurement can be further improved. This study and the ability to perform thorough parametric studies lay the groundwork for future advancements in antimatter research, paving the way for a deeper understanding of the fundamental forces governing our universe.

\begin{acknowledgments}
This research was supported by the Israel Science Foundation (ISF), Grant No. 1099/18.
\end{acknowledgments}

B.R. and G.F. contributed equally to this work.

\bibliography{Gravity_references.bib}

\end{document}


\preprint{APS/123-QED}

\title{A new technique to measure the gravitational mass of ultra-cold atoms in a quadrupole magnetic trap and its implications for anti-hydrogen studies}

\author{Boaz Raz}
\email{raz.boaz@weizmann.ac.il}
\author{Gavriel Fleurov}
\email{gavriel.fleurov@weizmann.ac.il}
\author{Roi Holtzman}
\author{Nir Davidson}
\affiliation{Department of Physics of Complex Systems, Weizmann Institute of Science, Rehovot 7610001, Israel}
\author{Eli Sarid}
\affiliation{Department of Physics, Ben-Gurion University of the Negev, P.O.B. 653, Beer-Sheva 84105, Israel}
\date{\today}

\maketitle
The supplementary material provides additional details and analysis for the experimental and simulation procedures described in the main text. It is ordered as follows. In Sec.~\ref{sec: experimental setup}, we present the experimental setup used for this experiment. In particular, we discuss the forced Radio-Frequency (RF) evaporation used to cool the atoms and the effect of the Majorana spin-flips and the lower boundary it places on the temperature of the cloud. In Sec.~\ref{sec: data analysis}, we describe the data analysis process. In Sec.~\ref{sec: simulation}, we present the simulation used to analyze the experimental data. We discuss the dynamics, the initial equilibrium distributions, and the verification of adiabatic cooling. In the final section, Sec.~\ref{sec: errors}, we present all the sources of statistical uncertainty and systematic error in the system.

\section{Experimental Setup}
\label{sec: experimental setup}
In this work, we use an existing cold atoms experimental setup \cite{edri_observation_2020,shkedrov_absence_2022}, following well-known \cite{davis_evaporative_1995,luiten_kinetic_1996,ketterle_boseeinstein_1996, muller_bose-einstein_2000,cohen-tannoudji_advances_2011} cooling and trapping techniques. As detailed below, forced RF evaporation is used to cool the atomic cloud to the various temperatures used in the experiment.

\subsection{State Preparation}
\label{sec: state prep}
The atomic cloud is prepared in a quadrupole magnetic trap in a mixture of $\ket{F=2,m_F=2}$ and $\ket{F=2,m_F=1}$ magnetic-trappable states, followed by forced RF evaporation process \cite{demarco_quantum_2001}, where atoms are transferred into an anti-trapped state ($m_F < 1$). The RF sweep employs a $\sim 10$ MHz sweeping RF signal, corresponding to the Zeeman splitting of the \isotope[87]{Rb} established in the magnetic trap. The atomic cloud is cooled by selectively removing the most energetic atoms, since they reach the farthest points in the trap and experience the largest magnetic fields, hence the largest Zeeman splitting. The $m_F = 2$ state has a magnetic moment twice as large as that of $m_F = 1$, so these atoms are lost from the trap before the $m_F = 1$ atoms since the sweep is from a high to a low frequency. The $m_F = 2$ then sympathetically cool the $m_F = 1$ atoms, so the $m_F = 1$ atoms are never directly evaporated. As a result, by the end of evaporation, the relative fraction of $m_F=1$ atoms is larger than it was at the beginning of the process.

\begin{figure}[ht]
    \centering
    \includegraphics[width = 0.85 \textwidth]{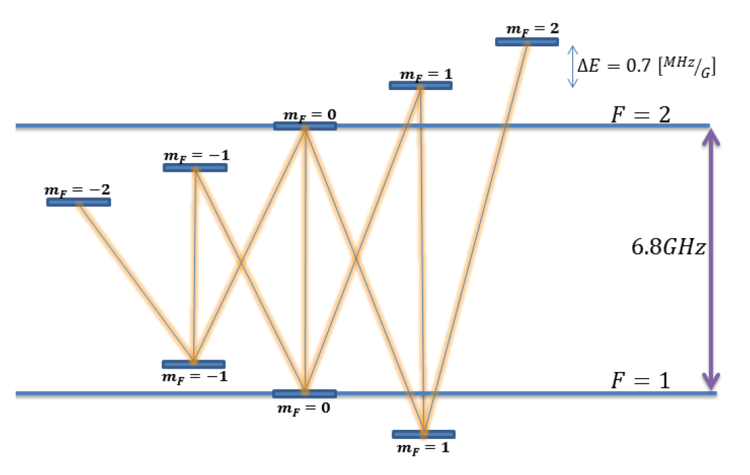}    
    \caption{\isotope[87]{Rb} hyperfine ground state transitions. $\Delta E$ is the linear Zeeman shift at small magnetic fields. Taken from \cite{daniel_coherence_2013}.}
    \label{fig: rb hyperfine ground  transitions}
\end{figure}

\begin{figure}[ht]
    \centering    
    \includegraphics[width = 0.85 \textwidth]{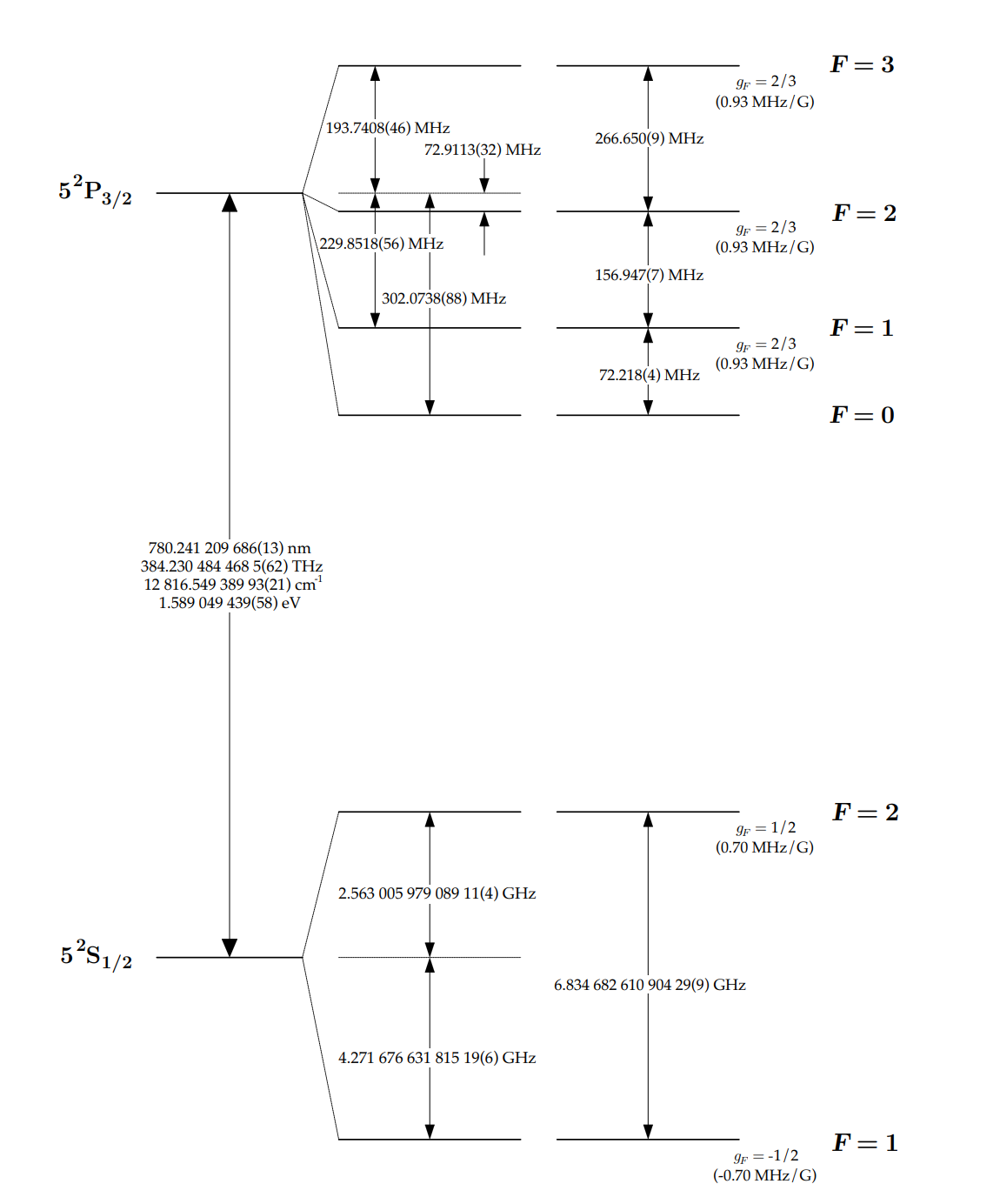}
    \caption{\isotope[87]{Rb}{} D2 transition hyperfine structure, with frequency splittings between the hyperfine energy levels. The approximate Land\'e $g_F$ factors for each level are also given, with the corresponding Zeeman splittings between adjacent magnetic sublevels. Taken from \cite{steck_rubidium_nodate}}
    \label{fig: rb hyperfine level diagram}
\end{figure}

The ratio of magnetic moments is also why our measurements are limited to the final magnetic field gradient  of $\beta_f = 2g$ (see main text for experimental protocol and definition of $\beta_f$). The $m_F = 1$ atoms survive up to that point and distort the counting of the atoms. Scanning $\beta_f$ from high to low values, we find two drops in the number of atoms at $\beta_f = g$ and $\beta_f = 2 g$, as shown in Fig. \ref{fig: wide magnetic gradient scan}. 

\subsection{Majorana Spin-Flips}
\label{sec: Majorana}

Majorana spin-flips \cite{petrich_stable_1995, davis_evaporative_1995,sukumar_spin-flip_1997,majorana_ettore_2006,lin_rapid_2009} losses occur when the relative rate of change of the magnetic field that the atom experiences $\abs{\flatfrac{\dot{B}}{B}}$ due to its velocity is faster than the Larmor frequency $\omega_L$. The atoms' magnetic moments can not follow these fast changes in the magnetic field, so they "flip" their magnetic state into an anti-trapped state.  This creates a "hole" in the trap, allowing the atoms to escape it. The size of the "hole" in the trap depends on the temperature of the atomic cloud. The center of the hole is located at the bottom of the trap, where the magnetic field is approaching zero, and the relative change in the magnetic field is diverging. This results in losing the coldest atoms from the trap since they spend the most time near the hole.

The loss rate due to Majorana spin-flips is proportional to $(\beta/T)^2$, where $T$ is the temperature of the cloud. During the experimental protocol, $\beta$ is lowered at a rate that allows adiabatic cooling; thus, the temperature follows $(T \propto \beta^{2/3})$ \cite{almog_high_2018}. Therefore, the overall loss rate is proportional to $(\beta/T)^2=\beta^{2/3}$. The higher the magnetic field gradient, the higher the loss rate. This means the loss rate is reduced when the magnetic field is lowered, so the survival probability is higher. This effect is opposite to the effect of gravity, where lowering $\beta$ causes the loss of more atoms. The probability of an atom being lost from the trap due to Majorana spin-flips is given by $P_{\text{Majorana}} = 1 - e^{-t\Gamma}$, where $\Gamma$ is the loss rate and $t$ is the time that the atoms are held in the trap. The Majorana losses become dominant for lower initial temperatures, reducing the atoms' lifetime in the trap and distorting the measured loss feature. Therefore, there is an optimal temperature for the experiment that balances Majorana's losses with the desire for lower temperatures, where the loss feature is narrower.

It should be noted that the exact formula for the loss rate due to Majorana spin-flips may depend on several factors, including the magnetic field gradient profile, the trap geometry, and the specific atomic species used. Therefore, the formula above should be considered a general scaling relation, and the specific parameters may need to be determined experimentally for each system. 

\begin{figure}
    \centering
    \includegraphics[width = \textwidth]{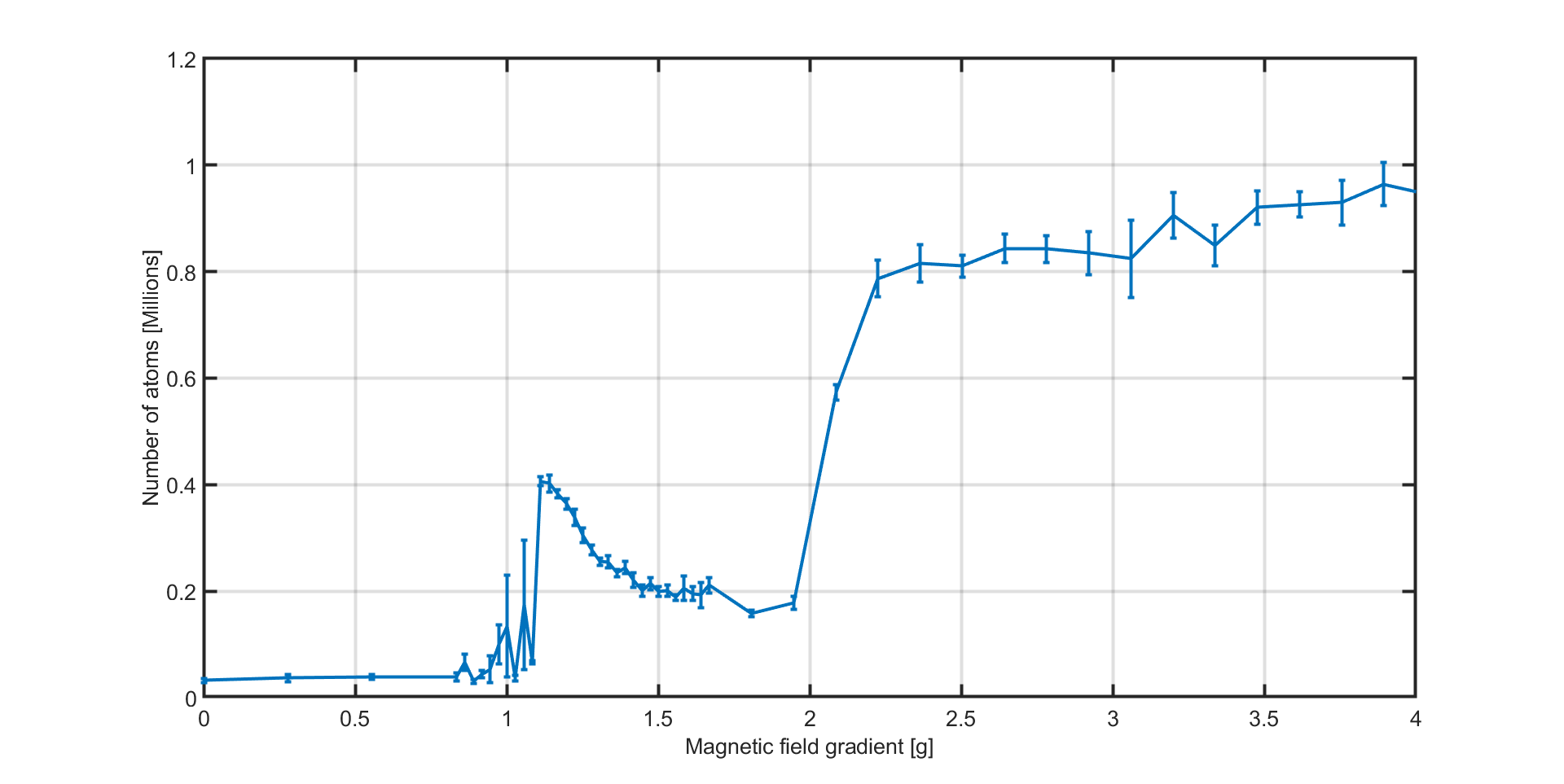}
    \caption{Two drops (from right to left) in the number of atoms as a function of $\beta_f$. The right drop corresponds to atoms in the $m_F = 1$ state, and the left drop to atoms in the $m_F = 2$ state. The rise to the left drop - moving from $\beta\simeq 1.5 g$ to $\beta\simeq 1.2 g$ is caused by the Majorana spin-flips; see text.}
    \label{fig: wide magnetic gradient scan}
\end{figure}

The effect of the Majorana spin-flips and the loss of atoms in different magnetic states are shown in Fig. \ref{fig: wide magnetic gradient scan}. When repeating the experiment with lower magnetic field gradients, the loss feature exhibits a rise in the number of atoms from $\beta_f = 1.5 g$ down to $\beta_f = 1.2 g$ before the drop due to gravity. Majorana losses become substantial, roughly around $T < 70 \mu K$ in our system.

\section{Data Analysis}
\label{sec: data analysis}
The experimental data comprises $\sim 11,000$ data points from 16 temperatures, $76$ values of $\beta_f$, and 9 repetitions of each value (see Fig. 1(c) in the main text). The system takes one data point every 35 seconds, leading to a total experiment time of almost a week. Over this extended period, the system is subject to several sources of long-term instabilities, such as laser power, frequency, and polarization fluctuations, drifts in ambient temperature and humidity in the lab, and magnetic field fluctuations. Most of these effects are mitigated by incorporating temperature measurements of the cloud between experiment points and monitoring the number of atoms loaded into the MOT at the initialization stages of the system by collecting fluorescence using a Photo-Multiplier Tube (PMT). The data is sampled wholly randomized to ensure that averaging cancels short-term fluctuations. The current in the coils is measured, and $\beta_f$ is determined during each point of the sequence.

\begin{figure}[ht]
    \centering
    \includegraphics[width = \textwidth]{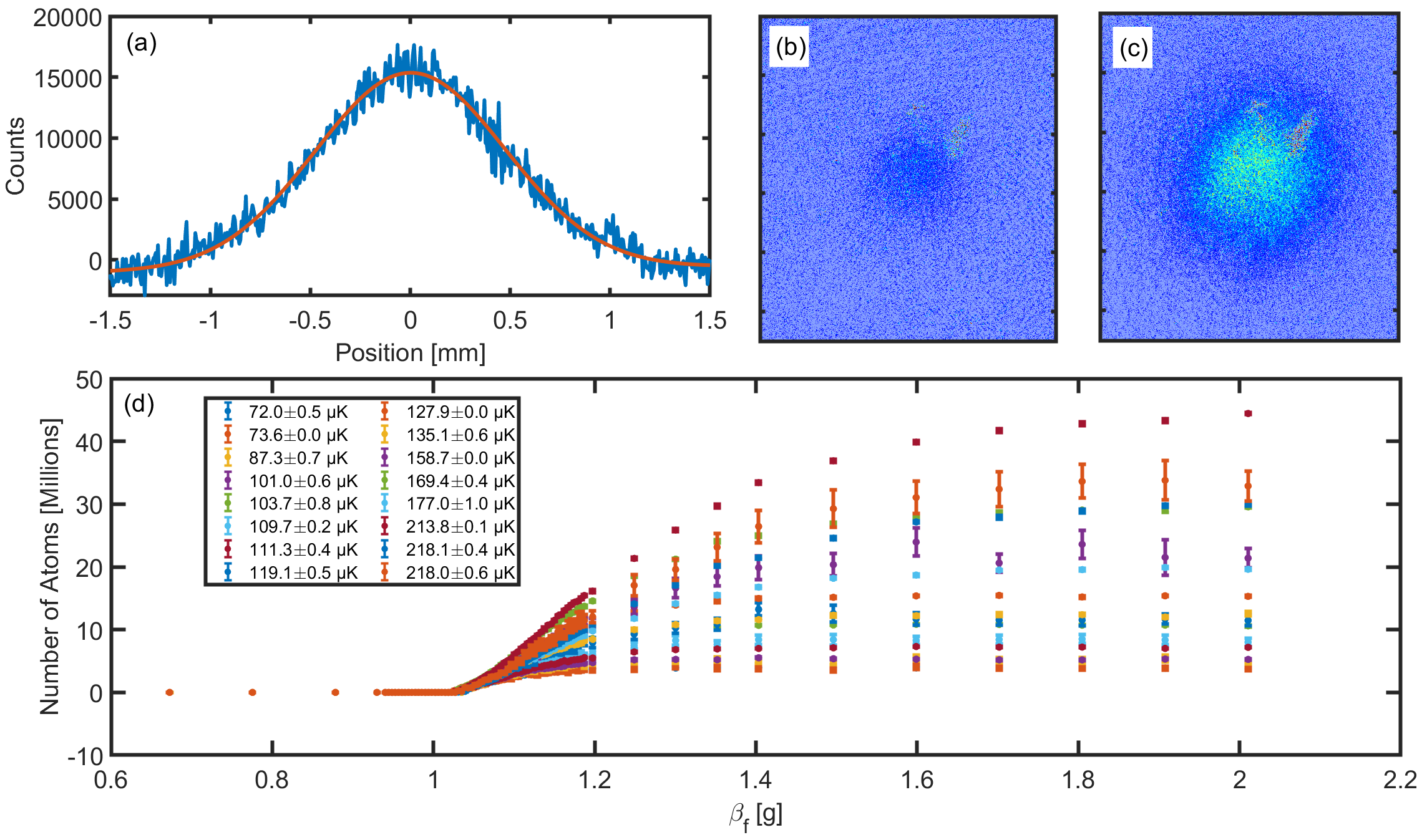}
    \caption{The data analysis process. (a) A Gaussian fit (red) to an atomic cloud's column density (blue) after ToF = 4 ms, horizontal axis centered around the fitted mean. (b-c) Absorption image of a cloud of 5.35 million atoms (b) and 0.8 million atoms (c). (d) Number of atoms left in the trap at the end of the experiment sequence for various temperatures.}
    \label{fig: data analysis}
\end{figure}

\begin{figure}[ht]
    \centering
    \includegraphics[width = \textwidth]{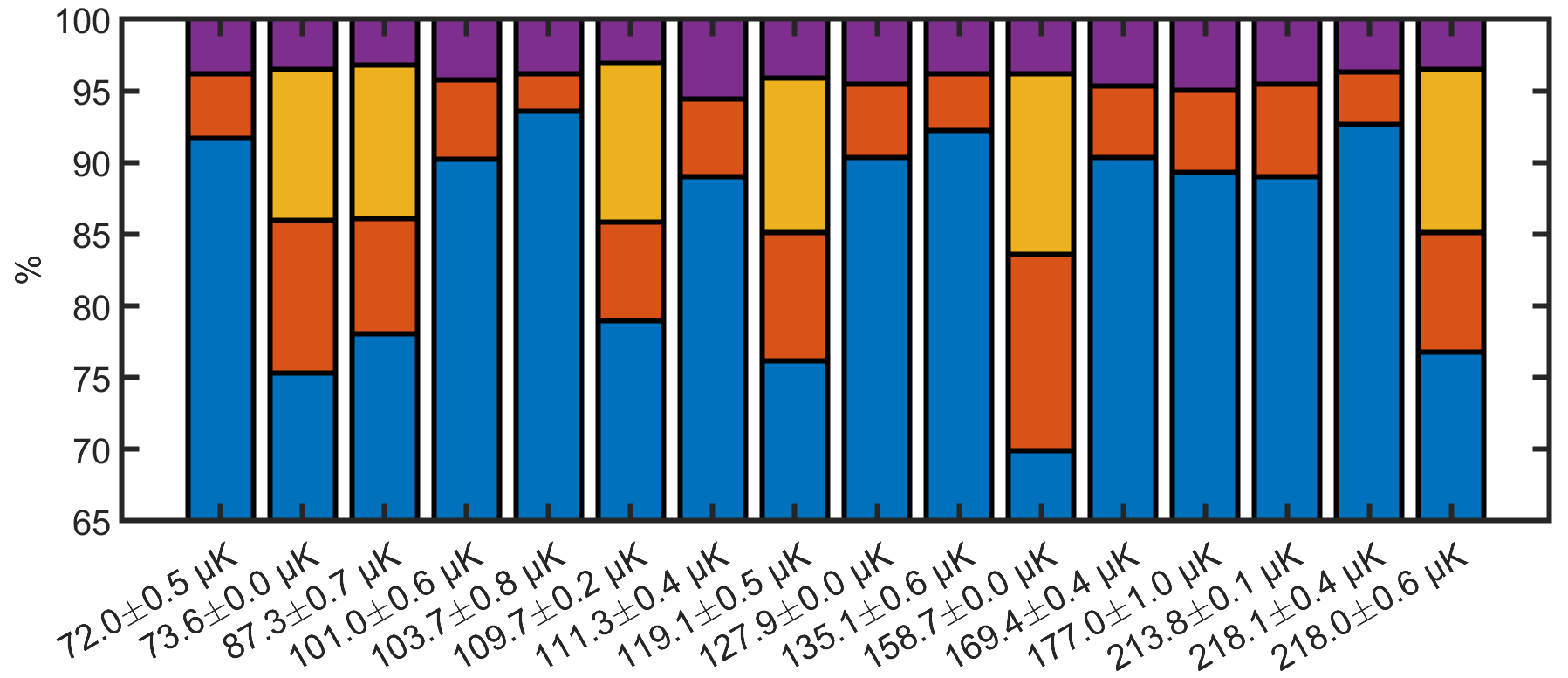}
    \caption{Experimental data usage for each temperature (Y-axis starts at 85\%). Data taken into the analysis (blue), data eliminated due to outliers after normalization (red), data eliminated due to low PMT (yellow), and data eliminated due to a mismatch in the electric current (purple).}
    \label{fig: experimental data usage}
\end{figure}

Furthermore, outliers more than three standard deviations from the mean of all other points with the same $\beta_f$ are eliminated. The percentage of points eliminated this way is marked in purple in Fig \ref{fig: experimental data usage}. Data points where the PMT shows significantly low numbers due to failure in loading into the MOT are disqualified. This step is crucial because, at lower magnetic field values, where only a few atoms are left, a nearly empty trap cannot be distinguished from a poorly loaded one - see Fig. \ref{fig: data analysis}(b-c). The data eliminated due to low PMT values are marked in yellow in Fig. \ref{fig: experimental data usage}.

The number of atoms is determined by absorption imaging of the atomic cloud. Four milliseconds after releasing the atoms from the trap, a resonant beam of light is shined onto the atoms and images them onto a CCD camera. The Optical Density (OD) is then computed \cite{ketterle_making_1999} from each image. An example image is shown in the inset in Fig. \ref{fig: data analysis}(a), and then the number of atoms is computed using the column densities of the OD.  The number of atoms is plotted as a function of $\beta_f$ in Fig. \ref{fig: data analysis}(d). Note the stark difference between the number of atoms at strong magnetic traps for different temperatures due to evaporation. 

The atom numbers are normalized to compute the survival probability. At each repetition for each temperature, the atom number is divided by the mean number of atoms of the five highest $\beta_f$ of that repetition - three highest for $T > 169$ µK. As discussed above, the measurements were carried out over a few days. Thus, each repetition is treated individually to avoid broadening the error bars due to long-term drifts. After normalization, outliers of more than three standard deviations from the mean are eliminated. These data points appear orange in Fig. \ref{fig: experimental data usage}(b).

\section{Simulation}
\label{sec: simulation}
In this section, we describe how the simulation of the evolution of the atoms cloud is performed. We assume that the atom cloud is (i) initiated in thermal equilibrium, namely, that the Boltzmann distribution gives the initial distribution, and (ii) that the atoms do not interact with each other during the protocol. These assumptions are justified since the collision rate reduces with the trapping strength, reaching less than 1 Hz during the experiment protocol, which takes 630 ms, so the contribution is negligible.

\subsection{The Dynamics of the System}
\label{sec: simulation dynamics}
The Hamiltonian of a single particle is given by
\begin{align}
    H(\vec{r}, \vec{p}, t) &= \frac{p^2}{2m} + U(\vec{r}, t) \\
    U(\vec{r}, t) & = g_F m_F \mu_B \beta(t) \sqrt{\frac{x^2+y^2}{4}+z^2} + m g z,    
    \label{eq:hamiltonian-1}
\end{align}
where $g_F$ is the hyperfine Landé g-factor, $m_F$ is the magnetic quantum number of the hyperfine level $F$, $\mu_B$ is Bohr's magneton, $\beta$ is the magnetic field gradient in the $z$ direction, which is chosen to point opposite to the direction of gravity, $g$ is the gravitational acceleration and $m$ is the atomic mass. This Hamiltonian determines the initial distribution
\begin{equation}
    \label{eq:boltzmann}
    \rho(\vec{r}, \vec{p}, T) = \frac{\exp \left( - H(\vec{r}, \vec{p}) / k_B T\right)}{\int d\vec{r} d\vec{p} \exp \left( - H(\vec{r}, \vec{p}) / k_B T\right)}. 
\end{equation}
Once an initial condition for a particle is chosen, the system evolves according to Hamilton's equations
\begin{align}
\label{eq: r dot}
\dot{\vec{r}} &= \frac{\partial H}{\partial \vec{p}} = \frac{\vec{p}}{m} \\
\label{eq: p dot}
\dot{\vec{p}} &= - \frac{\partial H}{\partial \vec{r}} = - g_F m_F \mu_B \beta(t) \frac{1}{\sqrt{\frac{x^2+y^2}{4}+z^2}} \left( \frac{x}{4}, \frac{y}{4}, z \right) - m g \hat{z}.
\end{align}

\subsection{Sampling Initial Conditions}
\label{sec:sample-init-cond}
\begin{figure}
    \centering
    \includegraphics[width = \textwidth]{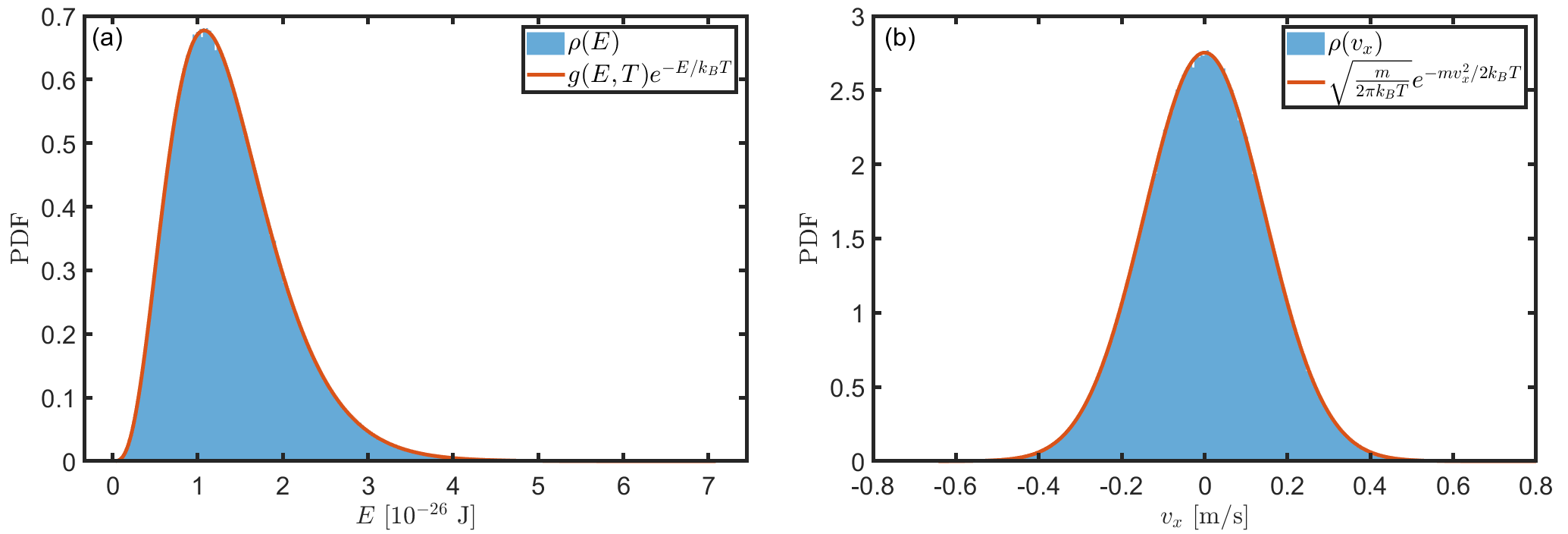}
    \caption{Initial distributions. (a) Histogram of $10^6$ instances of the sampled velocity $v_x$ using the Metropolis algorithm described in Sec.~\ref{sec:sample-init-cond}. The resulting histogram fits the expected Maxwell-Boltzmann distribution in Eq.~\eqref{eq:maxwell-boltzmann} with no fitting parameters. (b) Histogram of the total energy of $10^6$ initial conditions sampled using the Metropolis algorithm described in Sec.~\ref{sec:sample-init-cond}. The histogram fits the expected energy distribution in a quadrupole magnetic trap given in Eq.~\eqref{eq:rho-E} with no fitting parameters.}
    \label{fig: initial distributions}
\end{figure}
To properly sample the initial conditions from the Boltzmann distribution in Eq.~\eqref{eq:boltzmann}, we use the Metropolis algorithm \cite{metropolis_equation_1953}. The algorithm is performed as follows. A random uniform initial condition is chosen for both $\vec{r} = (x, y, z)$ and $\vec{p}$. First, a deviation $\delta \vec{r}$ is chosen uniformly. The new state $\vec{r}_{\rm new} = \vec{r} + \delta \vec{r}$ is proposed and accepted by the following condition. If the new state has lower energy from the current state, $H(\vec{r}_{\rm new}, \vec{p}) = E_{\rm new}< E_{\rm curr} = H(\vec{r}, \vec{p})$, then the new state is accepted. Otherwise, it is accepted with probability $\exp(- (E_{\rm new} - E_{\rm curr}) / k_B T)$. The same step is performed for a deviation in momentum $\delta \vec{p}$. These steps of proposing new coordinates and momenta are performed repeatedly 80,000 times. The resulting state $(\vec{r}, \vec{p})$ is then a state sampled from the Boltzmann distribution in Eq.~\eqref{eq:boltzmann}, because the dynamics of this algorithm satisfy detailed balance and it is ergodic. 

We have validated that the resulting distribution of this algorithm provides the Boltzmann distribution given in Eq.~\eqref{eq:boltzmann}.
The velocity distribution in each axis fits very well to a Gaussian with the proper variance, see Fig. \ref{fig: initial distributions}(b), given by $k_B T / m$, namely
\begin{equation}
\label{eq:maxwell-boltzmann}
    \rho(v_i, T) = \sqrt{\frac{m}{2 \pi T}} e^{- m v_i^2 / 2 k_B T}, \qquad i=x,y,z.
\end{equation}
Moreover, the  distribution of the total energy $E$ is given by \cite{luiten_kinetic_1996}
\begin{equation}
\label{eq:rho-E}
    \rho(E, T) = \frac{16}{105 \sqrt{\pi} (k_B T)^{9/2}} E^{7/2} e^{-E / k_B T}.
\end{equation}
Figure \ref{fig: initial distributions}(a) shows the sampled initial energies and the curve in Eq.~\eqref{eq:rho-E} with no fitting parameters, and they match very well.

We also check that the initial conditions satisfy the virial theorem to validate the simulation results further. The virial theorem is a relation between ensemble averages of the kinetic and potential energies. For a linear potential $V(r) \propto r^1$, it is
\begin{align}
    \label{eq: virial theorem}
    2 \expval{E_k} = \expval{V},
\end{align}
where $E_k$ is the kinetic energy. We compute the ensemble average energies of the atoms in the simulation at each time step and follow their dynamics. The atoms are evolved according to Eqs.~\eqref{eq: r dot} and \eqref{eq: p dot} with $\beta(t) = 20 \beta_*$, so their ensemble average should remain constant and satisfy Eq.~\eqref{eq: virial theorem}. See main text for the definition of $\beta_* = m g /g_F m_F \mu_B$. In conclusion, the virial theorem is satisfied within the simulation's statistical uncertainties, indicating that the simulations accurately capture the system's thermal equilibrium behavior. 

\subsection{Adiabatic Cooling}
\label{sec: simulation adiabatic cooling}
In a quadrupole trap, a slowly changing magnetic gradient $\beta$ guarantees adiabatic cooling \cite{almog_high_2018, kastberg_adiabatic_1995, gabrielse_adiabatic_2011}, as discussed above in Sec.~\ref{sec: Majorana}. The temperature will follow $\beta$ according to $T \propto \beta^{2/3}$. Note that this process requires no collisions, and the cloud remains in thermal equilibrium; therefore, we use the simulation to verify that the cloud's temperature changes adiabatically.

\begin{figure}[ht]
    \centering
    \includegraphics[width = 0.7\textwidth]{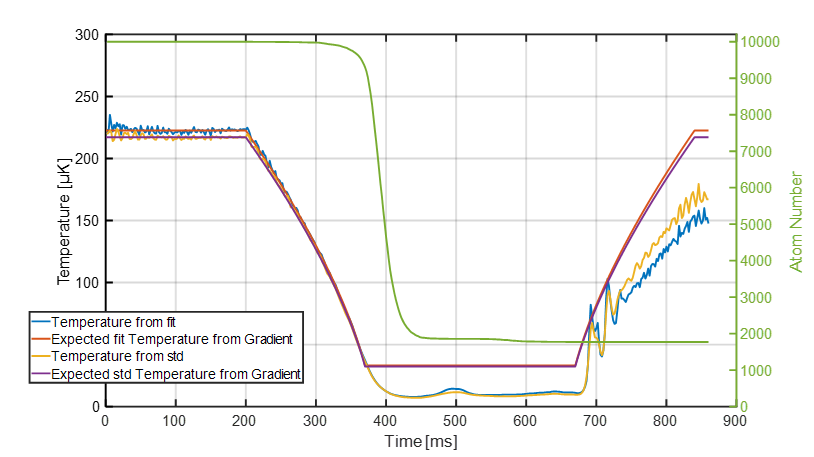}
    \caption{Validation of adiabatic cooling during the experimental protocol. The temperature in the simulation is measured using a Gaussian fit to the velocity distribution. The expected temperature according to adiabatic cooling is also plotted with the initial temperature taken as the mean of the initial plateau. Since atoms lost during the weakening of the trap remove heat from the system, the data of the rise back is not expected to follow the theoretical adiabatic heating.}
    \label{fig: adiabatic cooling}
\end{figure}

Fig. \ref{fig: adiabatic cooling} presents validation of adiabatic cooling based on the simulation. The cloud is initialized at $T = 220$~µK, and the simulation follows the experimental protocol. At each time step, the temperature of the cloud is calculated by comparing a Gaussian fit with the Maxwell-Boltzmann distribution. The mean along the initial plateau (Time $<$ 200 ms) is calculated and labeled as $T_0$, and the subsequent decrease of the temperature is calculated according to $T = T_0 \beta^{2/3}$. The "measured" temperature of the atoms in the simulation agrees with the expected temperatures according to adiabatic cooling until atoms are lost from the trap (green line). The temperatures are not relied upon after this point (Time > 370 ms).

\section{Statistical uncertainty and systematic errors}
\label{sec: errors}

\subsection{Magnetic Gradient Calibration}
\label{sec: mag field cal}
\begin{figure}[ht]
    \centering
    \includegraphics[width=\textwidth]{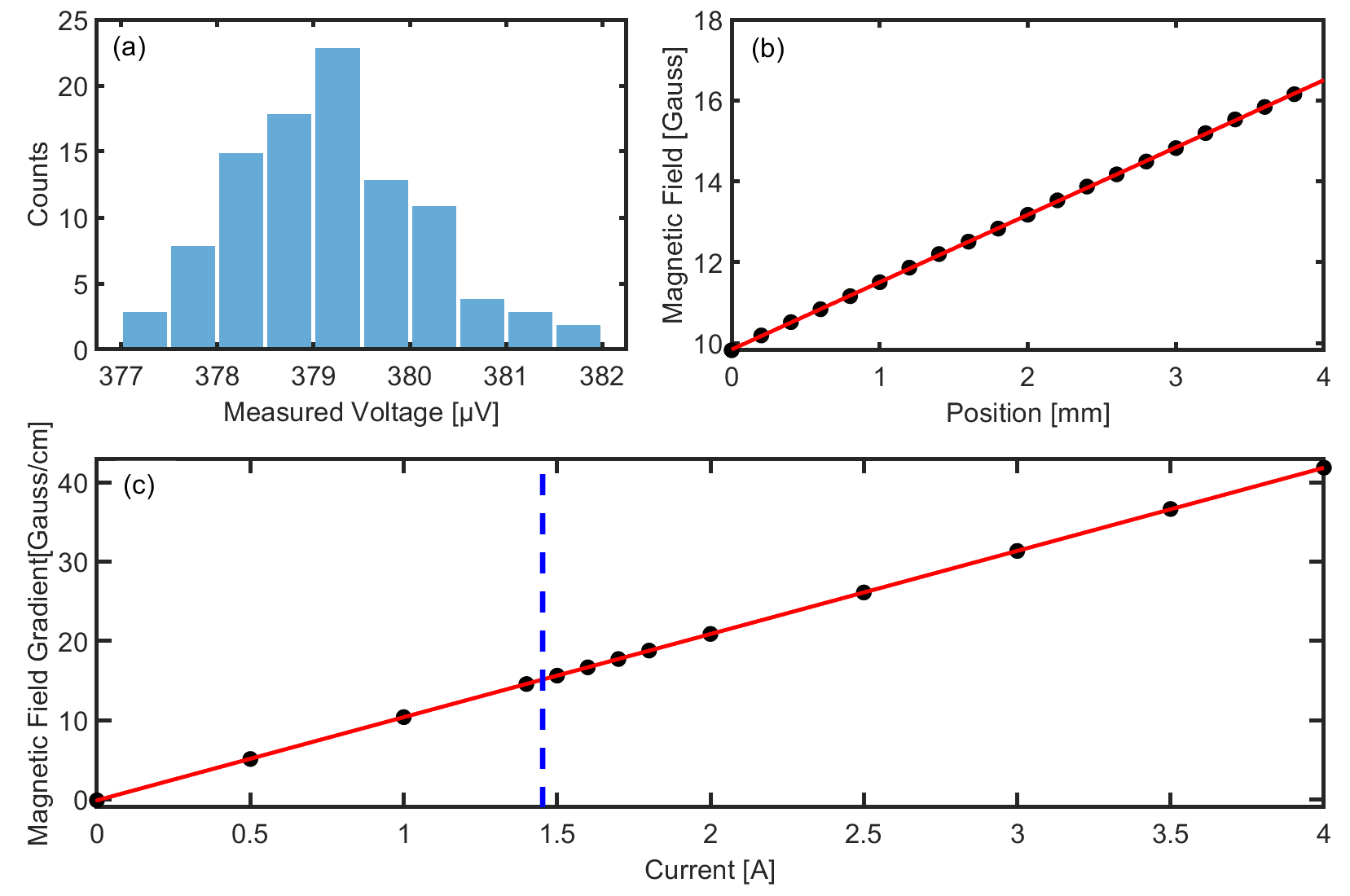}
    \caption{The calibration process of the magnetic gradient. (a) A histogram of the offset voltage on the \href{https://www.keysight.com/il/en/product/34470A/digital-multimeter-7-5-digit-truevolt-dmm.html}{Keysight 34470A} measurement. (b) An example of the magnetic field measurement for an electric current of 1.6A. Here $\beta = 8.330 \pm 0.028$ G/cm. (c) Calibration of the magnetic field gradient, the blue line is the value at which $\beta = \beta_*$. Error bars are smaller than the markers.}
    \label{fig: magnetic calibration}
\end{figure}

The system has two coils in an anti-Helmholtz configuration. Our workshop electrical engineers designed a current source with 2 ppm stability and noise. A real-time measurement of the current is taken during each shot, using a voltage reading of  \href{https://www.keysight.com/il/en/product/34470A/digital-multimeter-7-5-digit-truevolt-dmm.html}{Keysight 34470A}, measured on a stable resistor (Vishay VHP4ZT). We calibrated the current running through the coils and measured the inherent offset of the instrument, see Fig. \ref{fig: magnetic calibration}(a). The magnetic field gradient in our system was measured immediately after the experiment. We placed a \href{https://www.lakeshore.com/products/categories/overview/discontinued-products/discontinued-products/model-455-dsp-gaussmeter}{gaussmeter} (Lakeshore gaussmeter 455, $0.075\%$ accuracy) with an axial probe (HMNA-1904-VR, DC to 20 kHz, accuracy $\pm0.2\%$, stability $0.09$ G/K) on a computer-controlled \href{https://www.thorlabs.com/thorproduct.cfm?partnumber=ZST225B}{translation stage} (repeatability $<5$ µm). The magnetic field is measured along the gravitation axis between 5 and 9 mm above the atoms along the coils' symmetry axis using a custom holder to center the probe to the geometric center of the magnetic field coils. Although reaching the atoms' position is impossible due to the vacuum cell's physical constraints, the magnetic field remains linear in the entire measured region, making the extrapolation to the atom's position viable. This process was repeated for several values of currents. For example, Fig. \ref{fig: magnetic calibration}(b) shows the measurement at a set current of $1.6$ amps. A fitted linear line provides the evaluation of the gradient and uncertainty. The value of the gradient is taken at each current, including zero where the circuit was open, and fitted to find the linear relation between the magnetic field gradient and the applied current to be $\beta(I) = 10.490 \pm 0.005$ Gauss/(cm$\cdot$A) with a bias of $-0.134\pm0.007$ Gauss/cm.

Microwave (MW) spectroscopy can measure magnetic field gradients with high precision. Applying an MW field to the cold atoms can induce Rabi oscillations between atomic states, where the frequency depends on the magnetic field strength. An uncertainty in frequency of 1 kHz corresponds to a magnetic field resolution of about 1.4 mG.

The magnetic field gradient in our experiment can be measured with higher accuracy by probing the Rabi oscillation frequency of a cloud positioned with a spatial resolution of 5 µm along the desired axis. At 300 G/cm gradients, the magnetic field difference over 6 mm will be 180 G. The relative error in the measured gradient will be on the order of $10^{-5}$.

\subsection{Temperature Measurement}
\label{sec: temperature measurement}

\begin{figure}[ht]
    \centering
    \includegraphics[width=\textwidth]{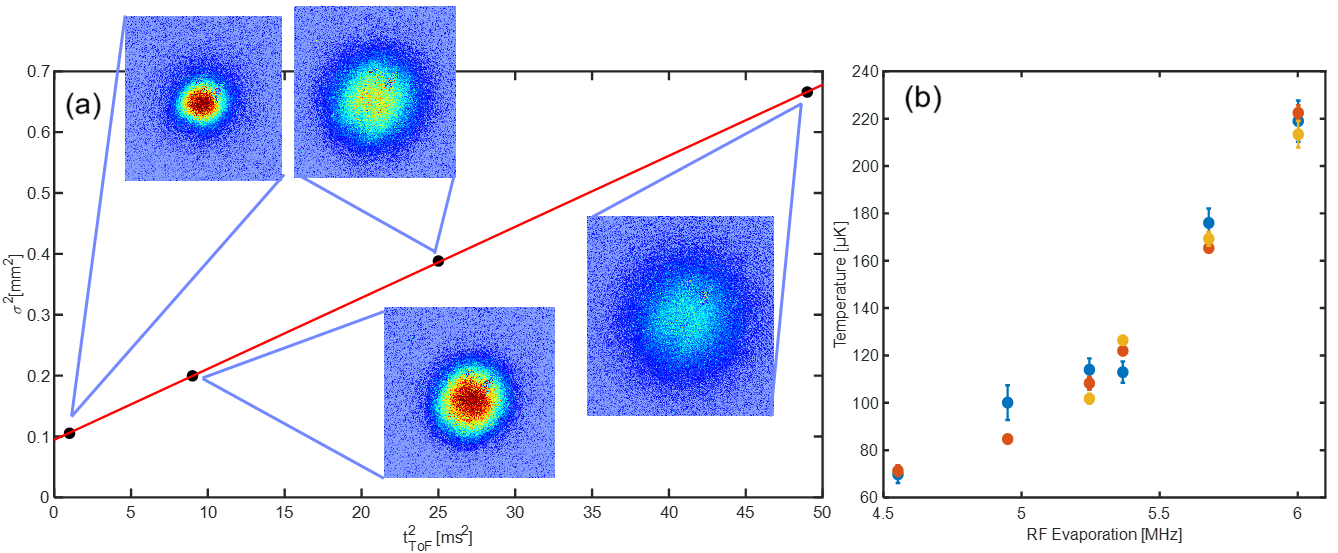}
    \caption{The temperature measurement in the system. (a) The width of the cloud squared is linearly fitted to the time of flight squared. A sample shot of each time of flight is presented. The markers are larger than their statistical errors. (b) Temperature measurements spread around their respective mean. Horizontal labels specify the RF evaporation cut-off frequency $\omega_{RF}$ and the combined uncertainty. Error bars represent statistical uncertainty. Different colors represent measurements taken at a $\sim 1$ day separation.}
    \label{fig: temperature measurement}
\end{figure}

The temperature of the atomic cloud is evaluated using the Time-of-Flight (ToF) technique \cite{anderson_observation_1995,ketterle_making_1999}, commonly used in cold atom experiments. The current through the magnetic coils is turned off abruptly ($< 1$ ms), releasing the atoms from the trap. The atoms are then imaged after different expansion times, and the absorption image is fitted with a Gaussian model. The width of the Gaussian, $\sigma$, is taken to be the size of the cloud. After long expansion times, when the cloud is much larger than the initial size in situ, $\sigma$ is correlated with the initial momentum distribution. The temperature in the trap is then given by $k_B T = m \sigma^2 /t^2$  \cite{ketterle_making_1999, regal_experimental_2006}. Here, $k_B$ is Boltzmann's constant, $m$ is the atomic mass, and $t$ is the ToF. The Gaussian width correlates with the trapped cloud's most probable velocity. In Fig. \ref{fig: temperature measurement}(a), a typical temperature measurement is plotted along with a linear fit. The temperature and its uncertainty are extracted from the slope. A typical absorption image of a cloud is shown for each data point.

The measurement campaign was conducted according to the following protocol. We chose six different RF evaporation cut-off frequencies $\omega_{RF}$ (see Fig. \ref{fig: temperature measurement}), calibrated to set the cloud at a range of temperatures between $70-230$ µK. The final magnetic gradient value $\beta_f$ was scanned over 76 different values in a fully randomized way to reduce biases due to fluctuating ambient magnetic fields. Each point was repeated 9 times. Each data point requires roughly 30 seconds, adding up to a day and a half of data collection. At the end of this session, a temperature measurement was conducted at each of the specified $\omega_{RF}$s. This entire session and the temperature measurements were repeated three times. The number of atoms and the resulting temperature of the cloud slightly drift on the scale of hours. Therefore, we found that for measurement sessions with the same $\omega_{RF}$ but on two different days, the resulting temperature differs on the order of 10 µK. But, as shown in Fig. \ref{fig: data analysis}(d), the atom numbers and temperatures in a single 9-repetition session are very stable.

\begin{figure}[ht]
    \centering
    \includegraphics[width=0.8\textwidth]{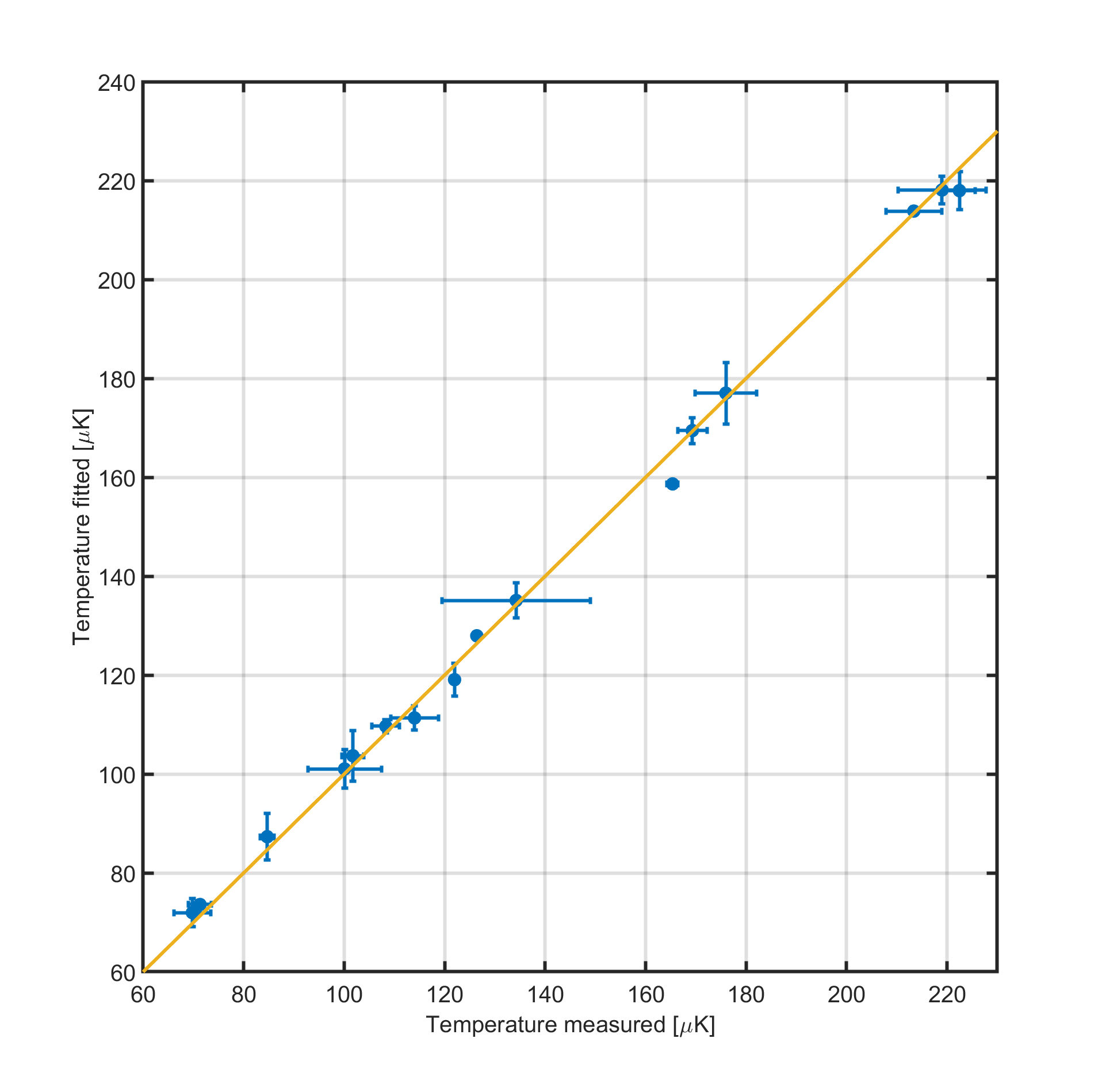}
    \caption{Comparison of Measured vs. Fitted Temperatures (blue), with error bars representing the $1\sigma$ uncertainty for each method. The yellow line indicates where the measured values equal the fitted values.}
    \label{fig:T_meas_vs_fit}
\end{figure}

This temperature measurement has a systematic error due to the extended duration (1 ms) of releasing atoms from the magnetic trap for imaging, leading to a small amount of adiabatic cooling before complete release. In Fig. \ref{fig:T_meas_vs_fit}, We compare the measured values' results with those found by fitting them to the simulation. 

\begin{figure}[ht]
    \centering
    \includegraphics[width=0.6\textwidth]{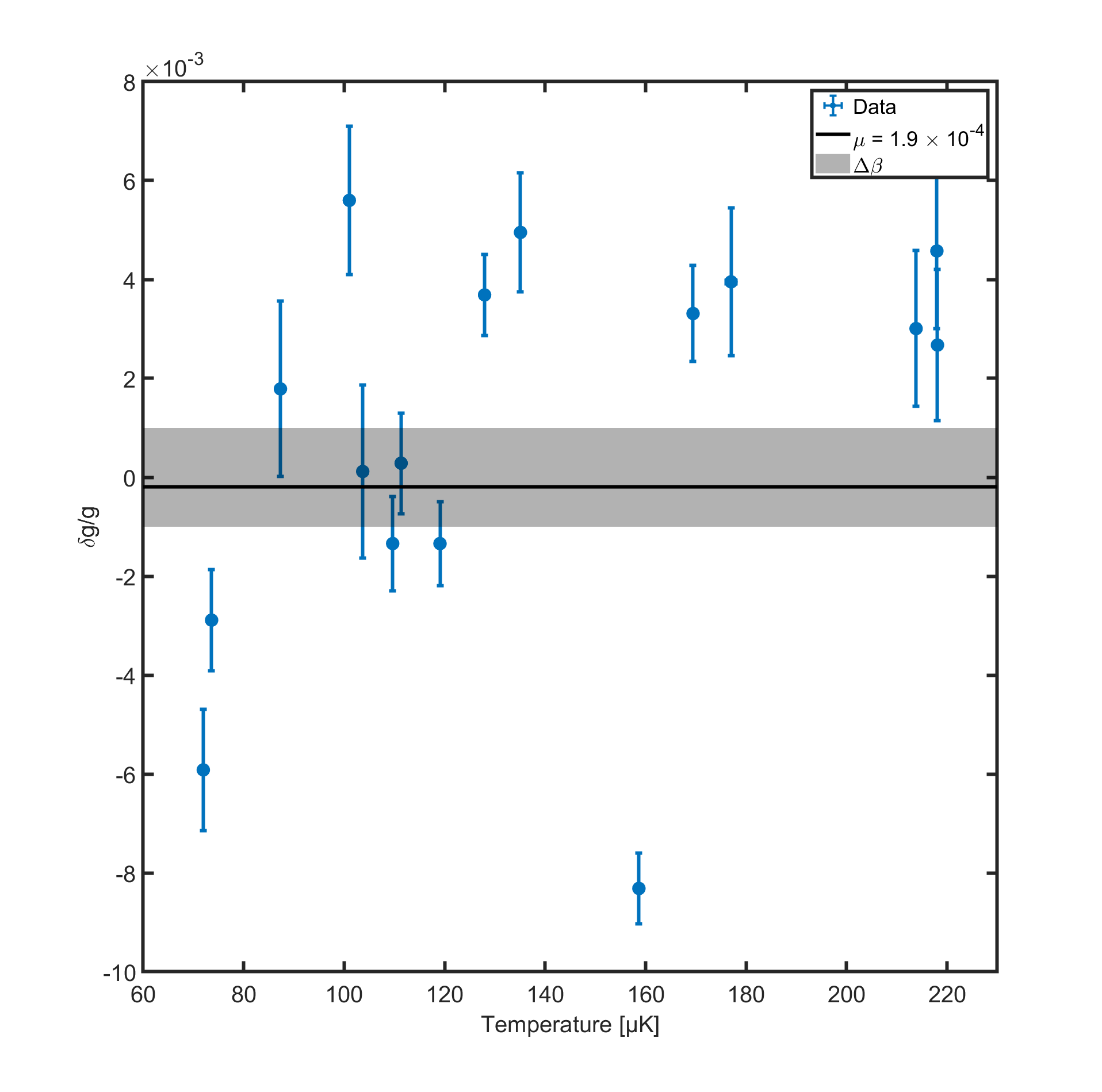}
    \caption{Relative error in the measured value of $g$ as a function of the measured temperatures with $1\sigma$ statistical error bars. The black line is the weighted mean $(-1.9\times 10^{-4}g)$ (with statistical uncertainty $(12\times 10^{-4}g)$). The gray shaded area is the uncertainty in the magnetic field gradient calibration $(5\times 10^{-4}g)$.}
    \label{fig:11}
\end{figure}

The results of our measurement for different temperatures are presented in Fig. \ref{fig:11}. The scatter of inferred values of $g$ across the 16 temperature points suggests a variation that extends beyond the error bars, indicating discrepancies that might not be accounted for by statistical uncertainty alone.

\begin{figure}[ht]
    \centering
    \includegraphics[width=0.9\textwidth]{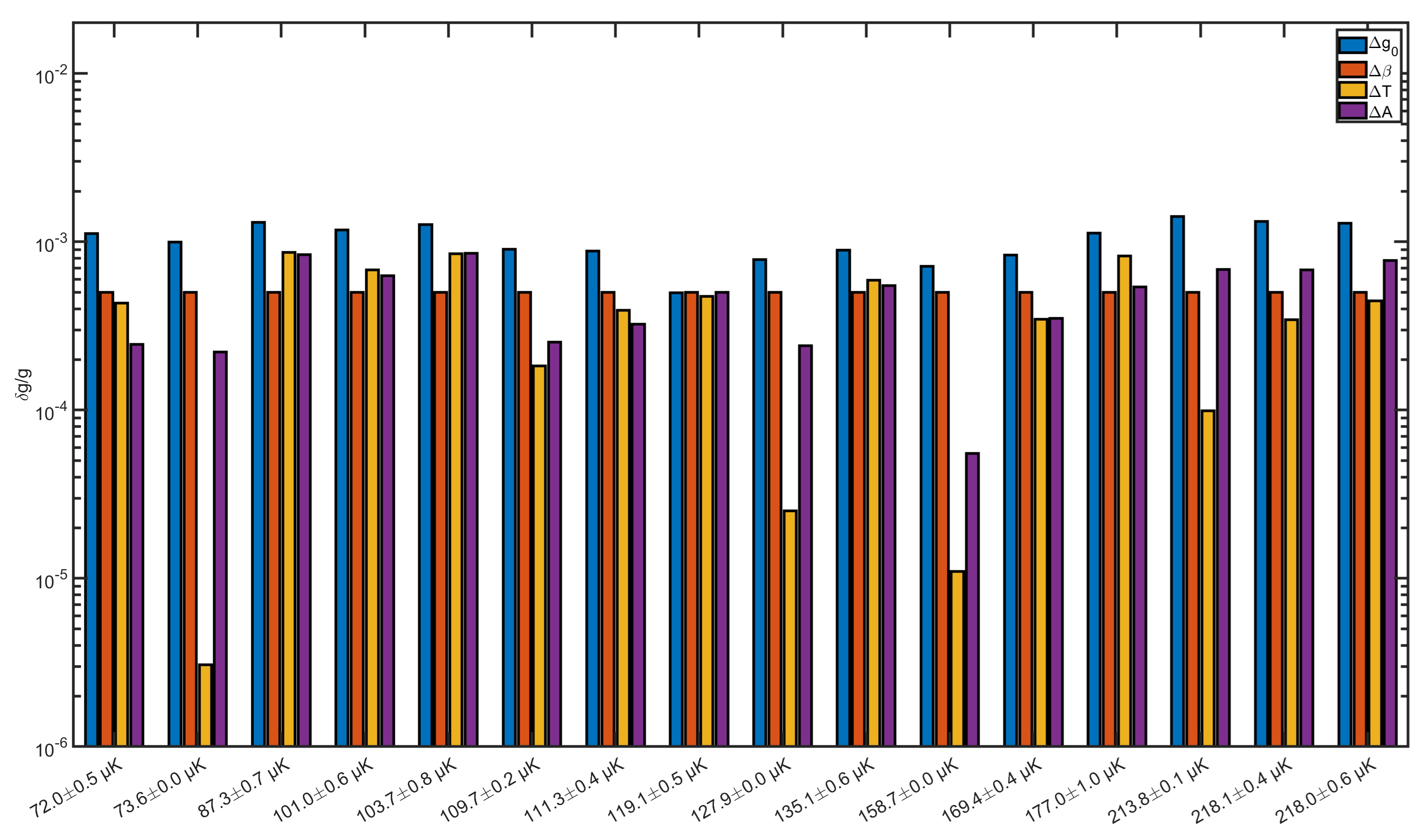}
    \caption{Relative Uncertainty in gravitational constant measurements across temperatures: blue for fitting, orange for magnetic field calibration, yellow for temperature fitting, and purple for normalization coefficient.}
    \label{fig:12}
\end{figure}

\subsection{Variation of $g$ with $A$ and $T$}
\label{sec: g derivatives}
To extract the effect of temperature $T$ and normalization factor $A$ on the final value of computed $g$, we scan these parameters in the simulation and extract the corresponding value of $g$ for each value of the scanned parameter by minimizing the error function $\epsilon(g,A)$ defined as:
\begin{align}
\epsilon(g,A,T) = \sum_j \frac{\qty(A \times P_j - f(\beta_{f,j};T,g))^2}{\sigma_{P,j}^2}
\end{align}
Where $P_j$ is the measured survival probability, $f(\beta_{f,j}; T,g)$ is the simulated survival probability, $\beta_{f,j}$ are the final magnetic gradient values of the experiment, and $\sigma_{P,j}$ is the one standard deviation uncertainty in the measured value of $P_j$.

\begin{figure}[ht]
    \centering
    \includegraphics[width = 0.8\textwidth]{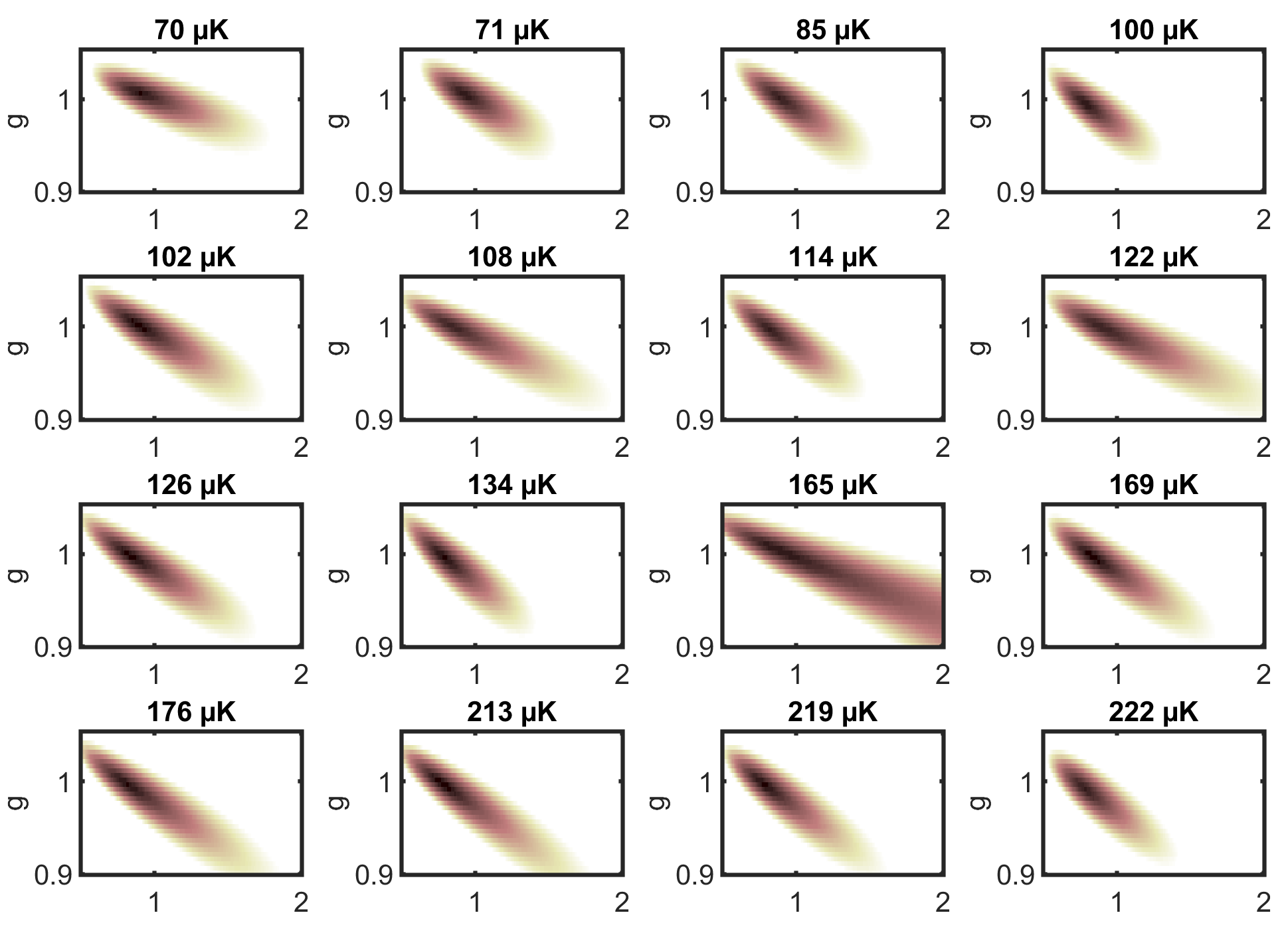}
    \caption{Error function $\epsilon(g,A,T = T_{exp})$ as a function of $g$ (vertical) and $A$ (horizontal) for all 16 measurement sets, at the measured temperature $T_{exp}$. The temperature at each set is written above each heatmap.}
    \label{fig: heat_map_ag}
\end{figure}

\begin{figure}[ht]
    \centering
    \includegraphics[width = 0.8\textwidth]{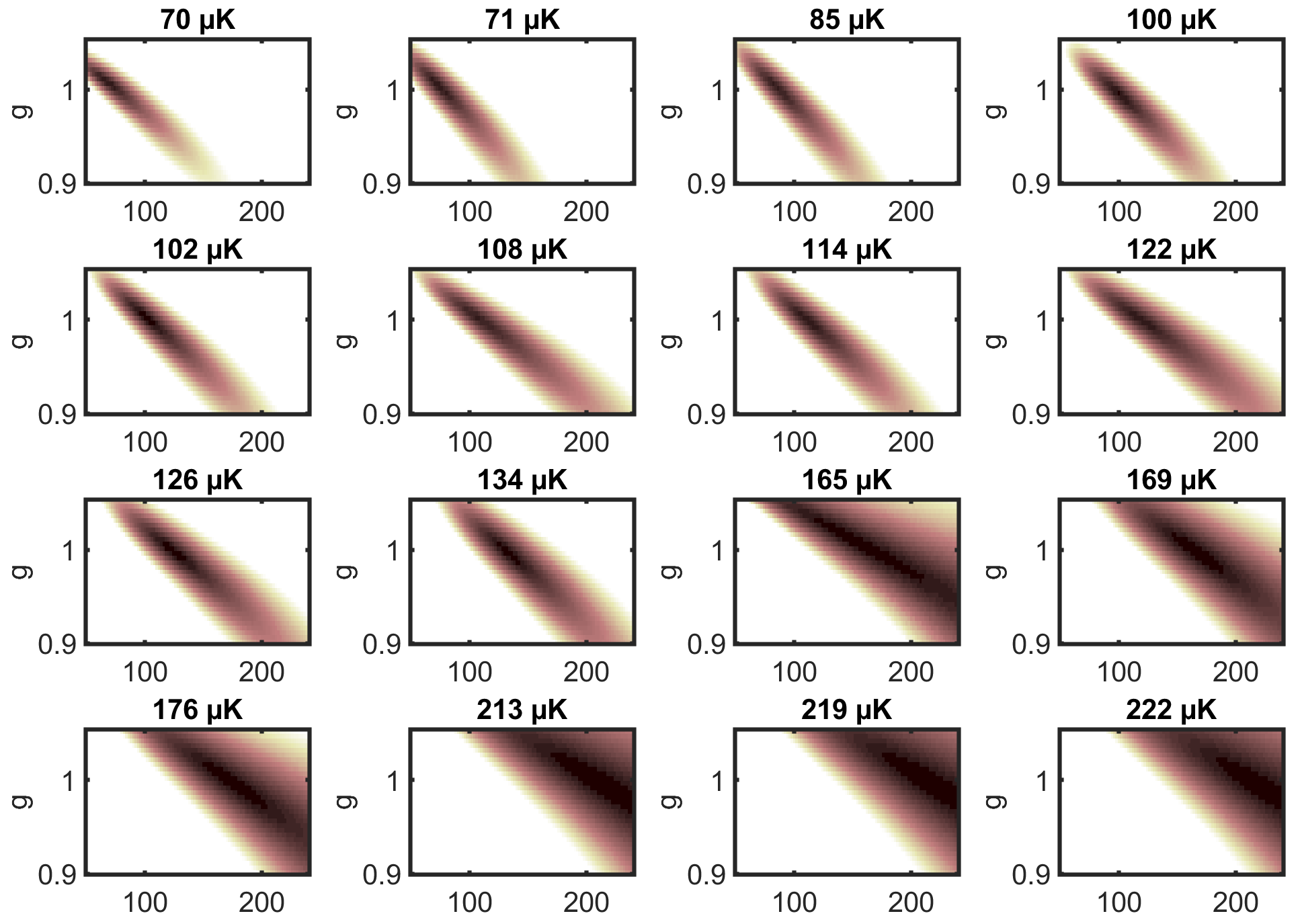}
    \caption{Error function $\epsilon(g,A = A_{opt},T)$ as a function of $g$ (vertical) and $T$ (horizontal) for all 16 measurement sets, at the optimal $A_{opt}$. The measured temperature at each set is written above each heatmap.}
    \label{fig: heat_map_tg}
\end{figure}

\begin{figure}[ht]
    \centering
    \includegraphics[width = \textwidth]{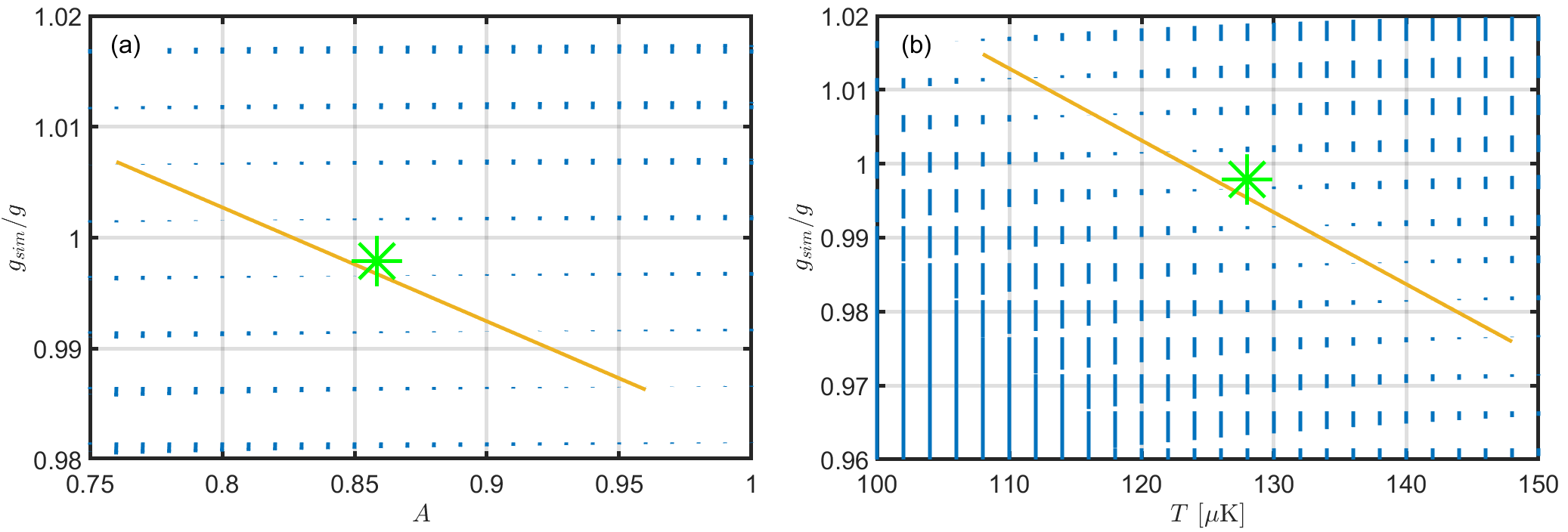}
    \caption{. (a) The gradient field of the error function $\epsilon(g,A)$ at a measured $T = 126.4 \pm 0.6$ µK. The green star is the position of the minimum of the error function, and the yellow line is the linear fit to the minimum of the gradient in its vicinity. The slope of the line is $\pdv{g}{A}$. (b) The gradient field of the error function $\epsilon(g,T)$ at a measured $T = 126.4 \pm 0.6$ µK and $A = 0.8585 \pm 0.0008$. The green star is the position of the minimum of the error function, and the yellow line is the linear fit to the minimum of the gradient in its vicinity. The slope of the line is $\pdv{g}{T}$.}
    \label{fig: gradient of A and T}
\end{figure}

Setting $T = T_\text{exp}$, where $T_\text{exp}$ is the experimentally measured temperature, We scan $A$ and $g$ to calculate the error function in the region of interest, see Fig. \ref{fig: heat_map_ag}. The minimum is found as depicted in the main text. The optimal $A$ for each measurement set was used to calculate the error function for a relevant range of $T$ and $g$, see Fig. \ref{fig: heat_map_tg}. Using $T$ from the independent temperature measurement as an input parameter, the optimal $g_{opt}$ was taken as the one found through the heatmap of $g-A$.

To estimate the error in $g_{opt}$ as a function of $A$ and $T$, a linear line was fitted to the direction of minimal change in $\epsilon(g,A,T)$ in a small region around $g_{opt}$. We find it by calculating the gradient fields $\pdv{g}{A}$ in Fig. \ref{fig: gradient of A and T}(a) and $\pdv{g}{T}$ in Fig. \ref{fig: gradient of A and T}(b), and finding a linear fit to the points of minimal gradient. The slopes of these lines are taken to be the values of the partial derivative presented in the main text.

\subsection{Atom counting}
\label{sec: atom counting}

Accurately counting the number of atoms in the trap is crucial for measuring gravitational acceleration. Several sources of error can contribute to inaccuracies in atom counting, as discussed in detail below.

The intensity of the imaging beam used for absorption imaging must be much lower than the saturation intensity $I_{sat}$. This is achieved using a low-power probe beam and carefully optimizing the polarization and detuning. Additionally, any residual magnetic fields in the imaging region can cause frequency shifts and cause an inhomogeneous broadening to the resonance linewidth, leading to errors in atom counting. These systematic effects are mitigated using common-mode rejection techniques, releasing the atoms from the same trap for each experimental sequence regardless of $\beta_f$.

The quantum efficiency of the camera used for imaging also introduces systematic errors in atom counting. We carefully calibrate the pixel size and account for dark image noise to correct for these effects. Furthermore, a reference image is acquired immediately after imaging the atoms to subtract any background noise or imaging imperfections in each shot.

Using the repumping light before and during imaging ensures that all atoms are in the same state before the absorption imaging. Aberrations in the imaging system and speckles in the imaging beam can also introduce systematic errors in atom counting. We carefully optimize the imaging system and beam parameters to minimize these effects.

Finally, atom counting can be challenging at low atom numbers, especially when approaching the detection limit. In this case, the relative counting uncertainty can become significant and contribute to errors in the measurement of gravitational acceleration. We carefully estimate the uncertainty in the number of atoms using statistical analysis of multiple measurements and propagate this uncertainty to the final measurement of $g$.
\FloatBarrier
\bibliography{Supplementary_ref.bib}